\documentclass[twocolumn]{aastex63}

\usepackage{ulem}
\usepackage{amsmath}
\usepackage{soul}

\shorttitle{Distances to WLM}
\shortauthors{Lee et al.}
\graphicspath{{./}{figures/}}

\begin{document}

\title{The Astrophysical Distance Scale III: \\Distance to the Local Group Galaxy WLM using Multi-Wavelength \\ Observations of the Tip of the Red Giant Branch, Cepheids, and JAGB Stars}

\author{Abigail~J.~Lee}\affil{Department of Astronomy \& Astrophysics, University of Chicago, 5640 South Ellis Avenue, Chicago, IL 60637}\email{abbyl@uchicago.edu}

\author{Wendy~L.~Freedman}\affil{Department of Astronomy \& Astrophysics, University of Chicago, 5640 South Ellis Avenue, Chicago, IL 60637}

\author{Barry~F.~Madore}\affil{Department of Astronomy \& Astrophysics, University of Chicago, 5640 South Ellis Avenue, Chicago, IL 60637}\affil{Observatories of the Carnegie Institution for Science 813 Santa Barbara St., Pasadena, CA~91101}

\author{Kayla~A.~Owens}\affil{Department of Astronomy \& Astrophysics, University of Chicago, 5640 South Ellis Avenue, Chicago, IL 60637}

\author{Andrew~J.~Monson}\affil{Department of Astronomy \& Astrophysics, Pennsylvania State University, 525 Davey Lab, University Park, PA 16802}

\author{Taylor~J.~Hoyt}\affil{Department of Astronomy \& Astrophysics, University of Chicago, 5640 South Ellis Avenue, Chicago, IL 60637}

\begin{abstract}
The local determination of the Hubble Constant sits at a crossroad. Current estimates of the local expansion rate of the Universe differ by about 1.7-$\sigma$, derived from the Cepheid and TRGB based calibrations, applied to type Ia supernovae. To help elucidate possible sources of systematic error causing the tension, we show in this study the recently developed distance indicator, the J-region Asymptotic Giant Branch (JAGB) method \citep{2020arXiv200510792M}, can serve as an independent cross-check and comparison with other local distance indicators. Furthermore, we make the case that the JAGB method has substantial potential as an independent, precise and accurate calibrator of type Ia supernovae for the determination of $H_0$. Using the Local Group galaxy, WLM we present distance comparisons between the JAGB method, a TRGB measurement at near-infrared ($JHK)$ wavelengths, a TRGB measurement in the optical $I$ band, and a multi-wavelength Cepheid period-luminosity relation determination. We find:

$$\mu_0 ~(JAGB)=24.97 \pm0.02 \text{ (stat)} \pm0.04 \text{ (sys) mag} $$
$$\mu_0 ~(TRGB_{NIR})=24.98 \pm0.04 \text{ (stat)} \pm0.07 \text{ (sys) mag}$$
$$\mu_0 ~(TRGB_{F814W})=24.93 \pm0.02 \text{ (stat)} \pm0.06 \text{ (sys) mag}$$ 
$$\mu_0 ~(Cepheids)=24.98 \pm0.03 \text{ (stat)} \pm0.04 \text{ (sys) mag}$$

\par\noindent
All four methods are in good agreement, confirming the local self-consistency of the four distance scales at the 3\% level, and 
adding confidence that the JAGB method is as accurate and as precise a distance indicator as either of the other three astrophysically-based methods.
\end{abstract}


\section{Introduction} \label{sec:intro}
Measuring both precise and accurate distances to galaxies persists as a challenging problem in astronomy. Attempts to discover, develop and refine new distance indicator techniques in the past two decades have only increased as the need for greater accuracy has grown. One goal of these endeavors, a robust calibration of type Ia supernovae, would provide a more secure determination of $H_0$ in the nearby universe, and a better understanding of the magnitude and significance of the current tension between local (late-time) values of $H_0$ and those derived from (early-time) cosmological modeling of cosmic microwave background (CMB) observations. Resolving the Hubble tension will require having systematic effects constrained at the 1-2\% percent-level (or better) in their individual accuracies. The two current front-runners for high-precision measurements of local distances are the Tip of the Red Giant Branch (TRGB) method used by the Carnegie-Chicago Hubble Program (CCHP) \citep{2019ApJ...882...34F,  2020ApJ...891...57F}, and  Cepheid Period-Luminosity based method (the Leavitt Law)  
used by the {\it Hubble Space Telescope Key Project} for the past 30 years \citep{2001ApJ...553...47F, 2012ApJ...758...24F} and more recently by the SHoES group \citep{2016ApJ...826...56R, 2019ApJ...876...85R}. Currently however, these two local (TRGB vs Cepheid) measurements of $H_0$ differ by almost 1.7-$\sigma$, and this tension continues. Is there an independent distance indicator, equal in precision and accuracy to the Cepheids and TRGB methods, that can serve as a cross-check to shed light on differences in systematics between the two more classical methods? We are now exploring that possibility.

The aim of this paper is to further develop an independent distance calibrator, the near-infrared {\it J-region} Asymptotic Giant Branch (JAGB) method, recently (re)introduced into the distance scale arena \citep{2020arXiv200510792M, 2020arXiv200510793F, 2020MNRAS.495.2858R}. 
JAGB stars are Thermally-Pulsating Asymptotic Giant Branch (TP-AGB) stars with considerable amounts of carbon in their atmospheres (i.e., carbon-to-oxygen ratios of C/O$ > 1$) \citep{2003agbs.conf.....H}. JAGB stars also undergo `dredge-up' events, where with each thermal pulse the convective envelope penetrates into deeper and deeper layers of the star's interior. During the third dredge-up phase (and beyond), the convective envelope of the star penetrates deep enough that it encounters carbon, produced by the He-burning shell, and brings that enriched material to the surface (see \cite{2003agbs.conf.....H} for a general review and \cite{2007A&A...469..239M, 2017ApJ...835...77M} for detailed descriptions of TP-AGB evolution).

For the younger, more massive AGB stars ($1.6 \times 10^8$ years), the temperature at the bottom of the convective envelope becomes so hot that during the third dredge-up phase, the recently-formed carbon is quickly converted into nitrogen before it can be convected to the surface. This places an upper limit on the luminosity of a JAGB star \citep{1973ApJ...185..209I, 1974ApJ...187..555S}.
For older, less massive AGB stars, the third dredge-up phase (and beyond) becomes ineffective in transporting carbon to the surface. This imposes a lower limit on the luminosities of stars entering the JAGB region \citep{1983ARA&A..21..271I}.
These theoretical limits are reflected in the observed properties JAGB stars, which are well-defined in their near-infrared (NIR) colors and luminosities. JAGB stars are  redder and quite distinct from the bluer oxygen-rich AGB stars, especially at near and mid-infrared wavelengths: a direct result of the large amounts of carbon in their atmospheres \citep{2007A&A...469..239M}. Thus, JAGB stars have been observed to have well-defined limits on their absolute magnitudes and color, making them  distinct and easily identified in color-magnitude space. For a more comprehensive review of the dredge-up theory for carbon stars, see \cite{2003agbs.conf.....H}. 

A review of carbon stars by \cite{2005A&A...434..657B} demonstrated that the mean luminosity of the distribution of JAGB stars is relatively constant from galaxy to galaxy, and preliminary tests carried out by \cite{2020arXiv200510793F} have also demonstrated the JAGB absolute magnitude has little, if any, dependence on metallicity or age.  Empirically, the mean of the JAGB luminosity distribution has all of the necessary attributes of an excellent distance indicator. 

Both the  TRGB and JAGB methods provide a standard candle. To compare the two methods, \cite{2020arXiv200510793F} measured and compiled distances to 14 galaxies, and found the TRGB and JAGB distances to be in good agreement. However, this study was based on previously-published data, not optimized for the detection of JAGB stars.  No studies to date have compared distance moduli using both the optical and near-infrared TRGB and the JAGB method derived from the same dataset, using self-consistent processing, calibration and analysis procedures.  Accordingly, in this study we undertake just such an in-depth study of the galaxy Wolf-Lundmark-Melotte  (WLM). 

WLM is a dwarf irregular galaxy located in the outer edges of the Local Group \citep{1909AN....183..187W,  1926MNRAS..86..636M}. We have measured its distance modulus using the TRGB in both the near-infrared $JHK$ bands and optical $I$ band, a multi-wavelength Cepheid period-luminosity relation determination, and the JAGB method in the $J$ band. We have obtained high-precision ground-based data, which we have calibrated using archival \textit{HST} observations for the optical and Two-Micron All-Sky Survey (2MASS) data \citep{2006AJ....131.1163S} for the near infrared. 

The first goal of this paper is to test the degree to which the JAGB method can be used as a cross check and comparison with other local distance indicators, such as Cepheids and the TRGB, in order to reveal potentially obfuscated systematics.  The second goal of this paper is to demonstrate that with future development and testing, using both ground-based and space-based data, the JAGB method is a distance indicator that is comparable to the TRGB and Cepheids in accuracy and precision, and can eventually become the basis for an independent determination of the Hubble Constant.

Our paper outline is as follows. In Section \ref{sec:data}, we describe the imaging datasets and the data-reduction methods used in this study. In Section \ref{sec:trgb}, we give a brief overview of the multi-wavelength application of the TRGB method, and then present our measurements of the distance modulus to WLM in the near-infrared and independently in the optical. In Section \ref{sec:jagb}, we explain the newly developed JAGB method, and then give our measured distance modulus. In Section \ref{sec:comparison}, we calculate a Cepheid-based distance modulus, as well as compare our TRGB distance moduli and JAGB distance modulus values with previous measurements for WLM. Finally, we state our conclusions and look to the future with the JAGB method in Section \ref{sec:summary}. We also provide supplementary information on our photometric procedure in Appendix \ref{App:qualitycut}.

\section{Data and Photometry} \label{sec:data}

\begin{figure*}
\gridline{\fig{"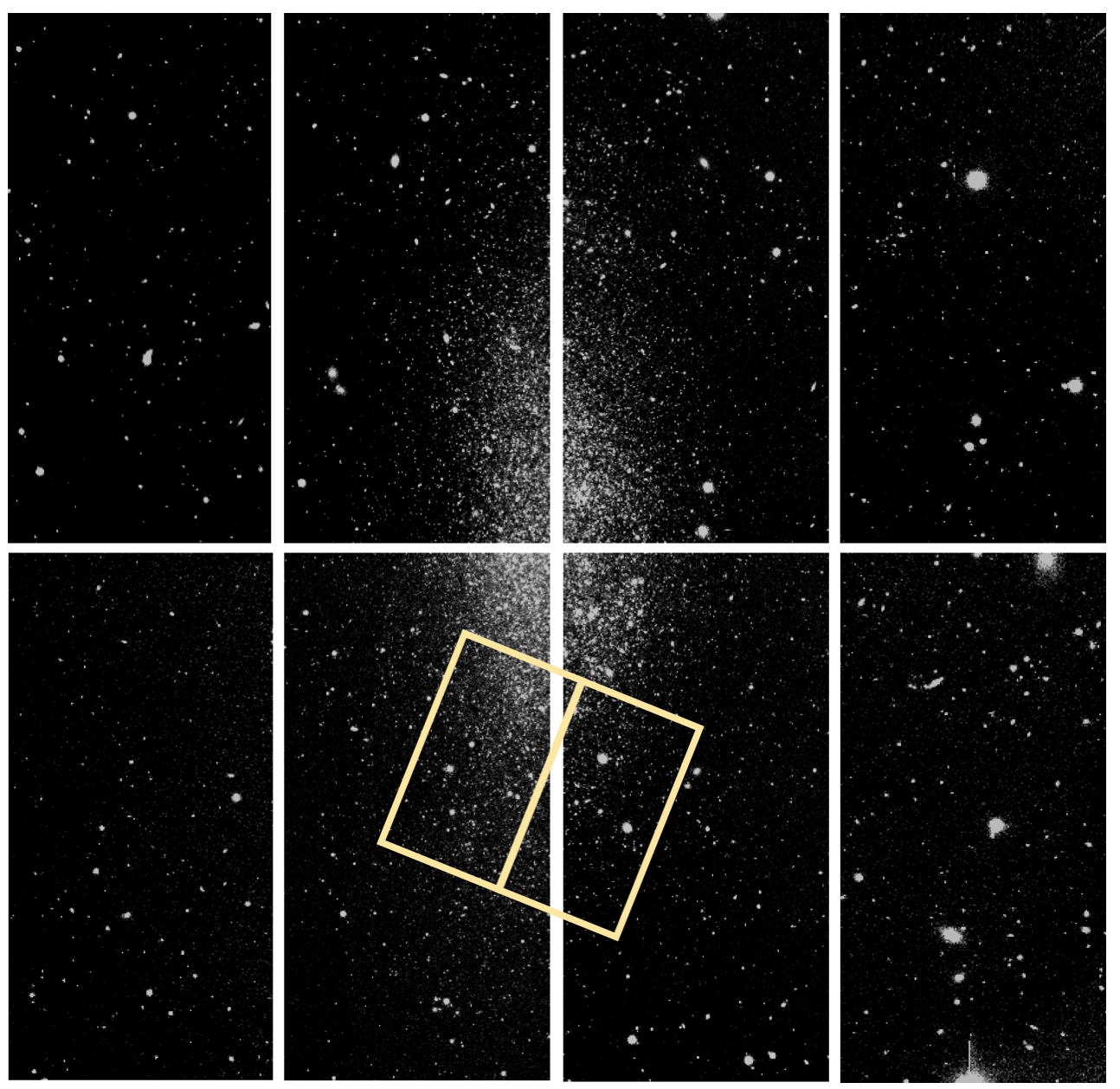"}{0.5\textwidth}{(a)}
          \fig{"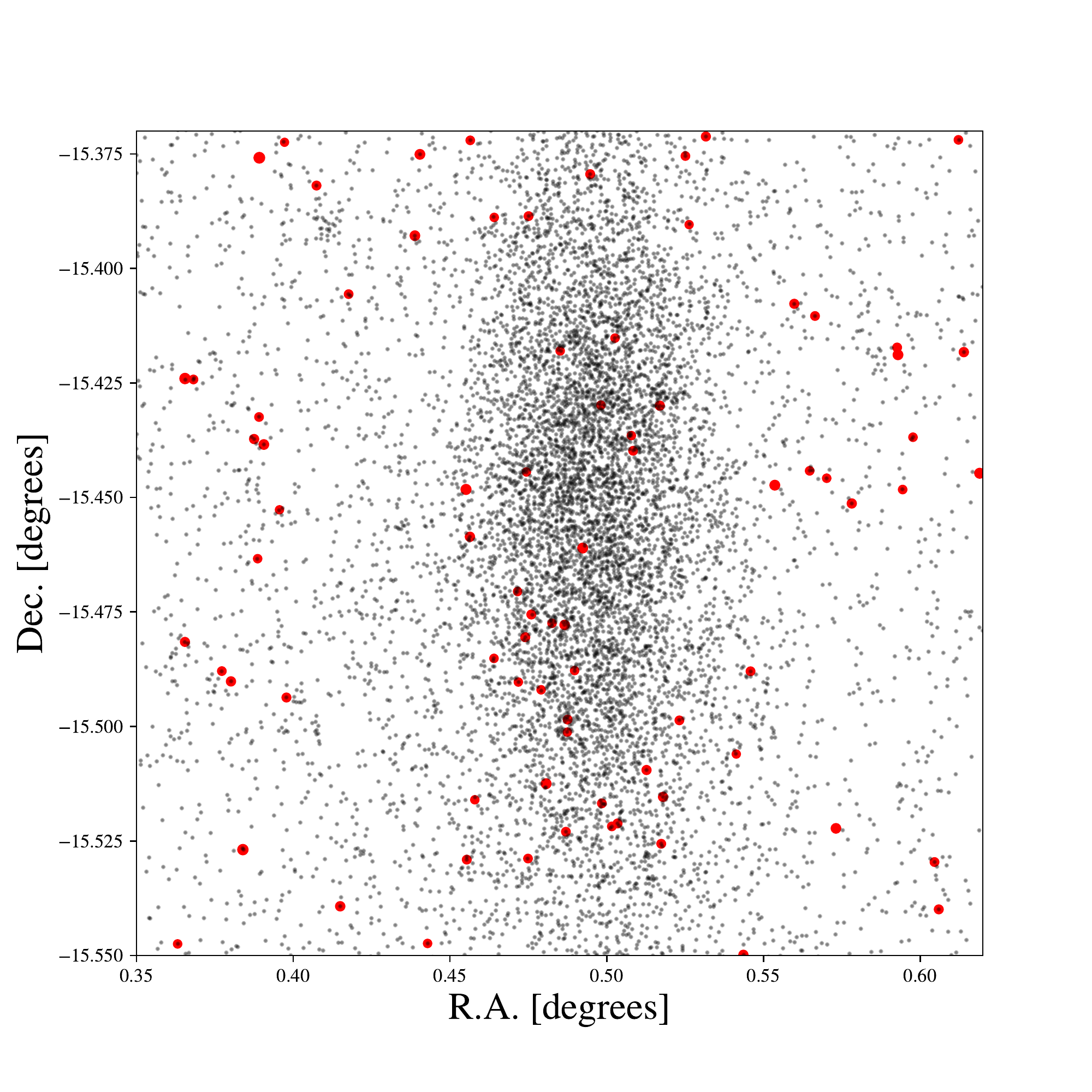"}{0.5\textwidth}{(b)}}
\caption{Images of the Local Group Galaxy WLM data. In Figure (a), $I$-band greyscale images for the eight IMACS chips are shown. The yellow rectangles show the locations of the two \textit{HST} ACS/WFC chips (F814W) used to calibrate our $I$-band ground-based photometry. Figure (b) shows the positions of all the queried 2MASS stars (red) overlaid onto the FourStar data (black) taken September 7, 2011. To give our IMACS and FourStar data right ascension / declination ($\alpha/\delta$) coordinates, we matched the data with HST/2MASS data that did have $\alpha/\delta$ positions.
\label{fig:f1}}
\end{figure*}

We now describe the five imaging datasets used in this study. A summary of the data can be found in Table \ref{tab:observations}. We first describe our process for obtaining final $VI$ photometry in Section \ref{sec:vi}, and then describe our process for obtaining $JHK$ photometry in Section \ref{subsec:jhk}.

\begin{deluxetable*}{cccccrc}
\tablenum{1}
\tablecaption{A Summary of Observations used in this work\label{tab:observations}}
\tablehead{
\colhead{Dates} & 
\colhead{Instrument} & 
\colhead{Filter(s)} & 
\colhead{$\alpha$} & 
\colhead{$\delta$} & 
\colhead{Exp. (sec)} &
\colhead{Target}
}
\startdata
2019-11-16 & IMACS & $VI$ & 00:01:58.0 & -15:27:39.4 & 1,800 sec & $TRGB_{F814W}$\\
2015-07-17,18,19 & ACS/WFC & F814W & 00:01:57.3 & -15:31:22.7 & 34,320 sec & $I$ calibration \\
2011-09-07 & FourStar & $JHK$ & 00:01:57:0 & -15:27:36.0 & $625$ sec & $TRGB_{JHK}$, $JAGB$\\
2011-10-05 & FourStar & $JHK$ & 00:01:57:0 & -15:27:36.0 & $1,083$ sec & $JAGB$\\
2019-11-13 & FourStar & $JHK$ &  00:01:57:0 & -15:27:36.0 & $39$ sec & $JHK$ Calibration
\enddata
\end{deluxetable*}

\subsection{VI Photometry}\label{sec:vi}

\subsubsection{Magellan-Baade Telescope: IMACS} \label{subsubsec:imacs}
Ground-based $VI$ data of WLM were obtained at the  \emph{Inamori-Magellan Areal Camera and Spectrograph} \citep[IMACS]{dressler_2011} on the $6.5$m Magellan-Baade telescope using the f/4 imaging mode at Las Campanas Observatory on November 16, 2019. The field has a dimension of $15.4\arcmin \times 15.4 \arcmin$ at a scale of 0.111 arcsec/pixel. 
The raw CCD data were then reduced using standard procedures:  the overscan region  was first subtracted by column from the raw science image; then the data were flat-fielded. 

\subsubsection{\textit{HST} Archival Data}
We utilized $HST$ archival data in order to calibrate the $I$-band ground-based photometry to the Vegamag magnitude system. We found one \textit{HST/ACS} dataset of WLM in the \textit{HST} archives that overlapped with the IMACS chips: \emph{Completing the Census of Isolated Dwarf Galaxy Star Formation Histories} \citep[PID: GO13768, PI: Weisz]{weiszprop}. The goal of this dataset was to measure the star formation histories of isolated dwarf galaxies by obtaining deep optical data. This dataset contained F814W data (the $HST$ filter corresponding to the Kron-Cousins $I$ band) which we used to tie our $I$-band ground-based photometry onto the \textit{HST} flight magnitude system. We note that color was used only to obtain our $VI$ color magnitude and visually check that the tip detection was reliable; the $V$-band are not used in the actual measurement of the TRGB in the $I$-band luminosity function.  

The $HST$ dataset contained 13 pointings, each 2,640 seconds, for a total exposure time of 34,320 seconds. The location of this pointing relative to the 8 IMACS chips is shown in Figure \ref{fig:f1}. The STScI processed and charge-transfer-efficiency-corrected and flat-fielded individual frames (the \textit{.flc} data type) were pulled from the Mikulski Archive for Space Telescopes (MAST).  We corrected for pixel-to-pixel geometric distortions caused by the optical design of the ACS camera by multiplying each image by its respective Pixel Area Map, which was pulled from the \textsc{stsci.skypac} module.\footnote{\url{https://www.stsci.edu/hst/instrumentation/acs/data-analysis/pixel-area-maps}}

\subsubsection{$VI$ Instrumental Photometry} \label{subsec:phot}
To obtain instrumental magnitudes from our imaging datasets, we
followed the standard photometry procedure
for obtaining PSF-photometry outlined in
the \textsc{daophot-II} User Manual
\citep{DAOPHOTIIMAN2000}. The procedures
were identical for both the \textit{HST} archival data and the IMACS data with the
exception that for the IMACS data, 
empirical point-spread functions were
manually created, and for the \textit{HST}
data, theoretical model-based TinyTim
point-spread functions
\citep{2011SPIE.8127E..0JK} were
generated. TinyTim PSFs were used here to
be consistent with the $I$-band absolute
magnitude TRGB calibration undertaken by the CCHP
\citep{2019ApJ...882...34F}, which
also used TinyTim PSFs. 

The basic photometry procedure is as follows: for a given band, photometry was first performed on each frame using the \textsc{daophot} suite, resulting in instrumental magnitudes for each frame.  Then, for a given chip, the images were aligned using \textsc{daomatch/daomaster}, and then a medianed image was created with \textsc{montage2}. Subsequently, the medianed image was photometered to create a master source list, and then all of the individual frames were simultaneously photometered using \textsc{allframe} and the master source list. Finally, all of the final photometry files were matched using \textsc{daomaster} again to give final instrumental magnitudes for a given chip. This process was repeated for all chips and bands.

\subsubsection{Calibrating \textit{HST} Archival Data}

In order to calibrate the instrumental \textit{HST} magnitudes onto the Vegamag magnitude system, we followed the procedure outlined in \citet{2005PASP..117.1049S}. First, the flight magnitude zero point was obtained from the STScI on-line calculator\footnote{\url{https://acszeropoints.stsci.edu/}}. For our \textit{HST} observations taken July 17, 2015, the ACS zeropoint reported was $ZP_{F814W}=25.517$ mag.
Second, the infinite aperture correction, the correction from a $0.5\arcsec$-to-infinity aperture, was obtained from \cite{2016AJ....152...60B}, determined to be 0.098 mag. We adopted a $\pm0.02$ mag error for both the zeropoint and infinite aperture corrected consistent with the CCHP program \citep{2019ApJ...885..141B}.  Third, we needed the correction from the PSF magnitudes to the $0.5\arcsec$ aperture. This correction was obtained by first identifying bright and isolated stars in the medianed image, and then re-locating them all in 26 frames. We computed growth curves for each star using \textsc{daogrow} \citep{1990PASP..102..932S}, and eliminated the few stars where the growth curve showed evidence that the star's flux was being contaminated from neighboring stars. 
Now left with 30 stars for a given frame, the aperture magnitude at  $0.5\arcsec$ minus the PSF magnitude was calculated for all the stars. After eliminating outliers more than 2-$\sigma$ from the median value, we averaged all the stars' individual aperture corrections to give the final aperture correction for a given frame. Aperture corrections were then applied on a frame-by-frame basis. The average aperture correction across all frames was $-0.107 \pm0.009$ mag.

\subsubsection{Calibration of $I$-band Photometry onto HST Flight Magnitude System}\label{subsubsec:calibrationVI}
\begin{figure}
\gridline{\includegraphics[width=\columnwidth]{"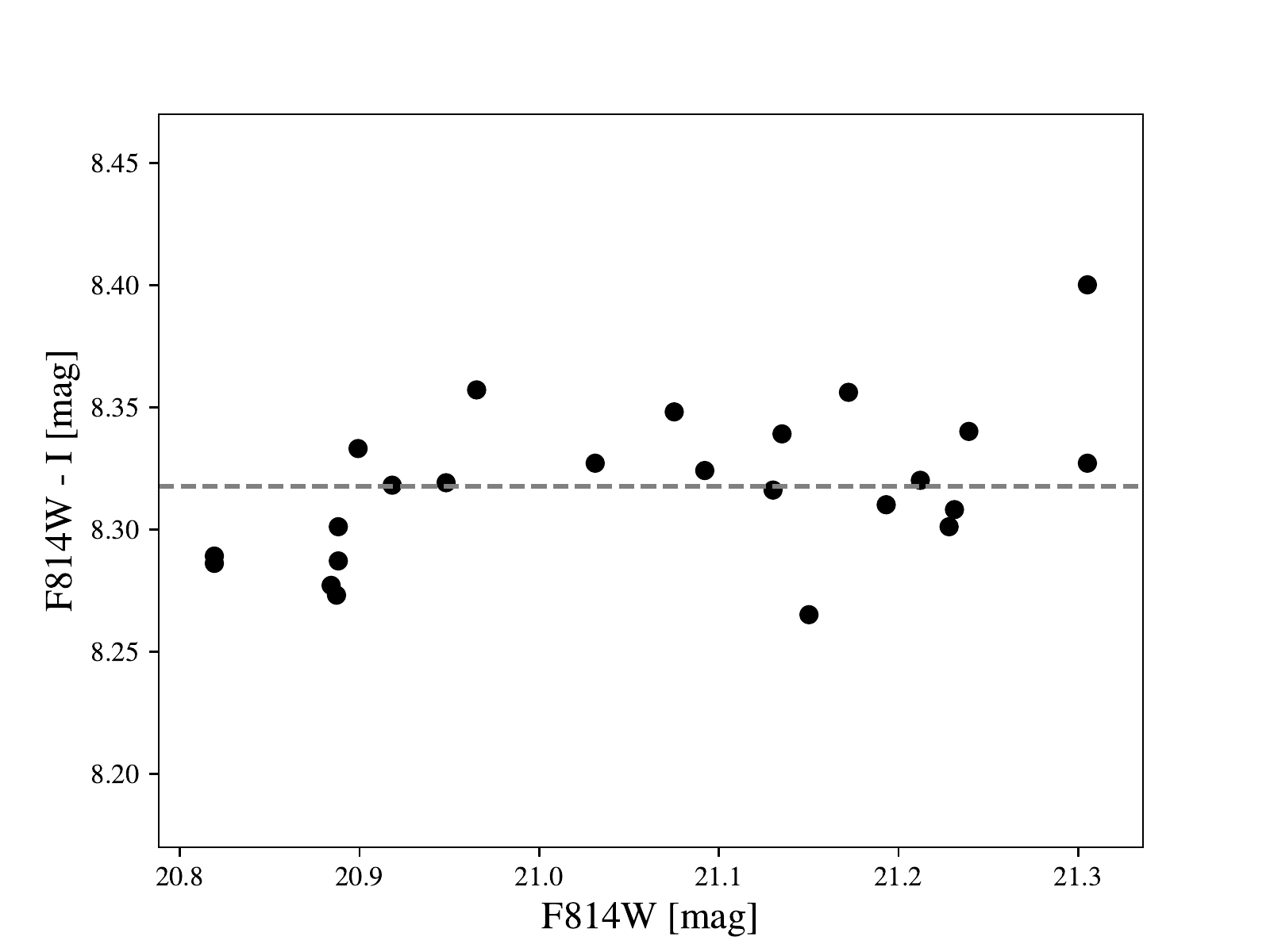"}{(a)}}
\gridline{\includegraphics[width=\columnwidth]{"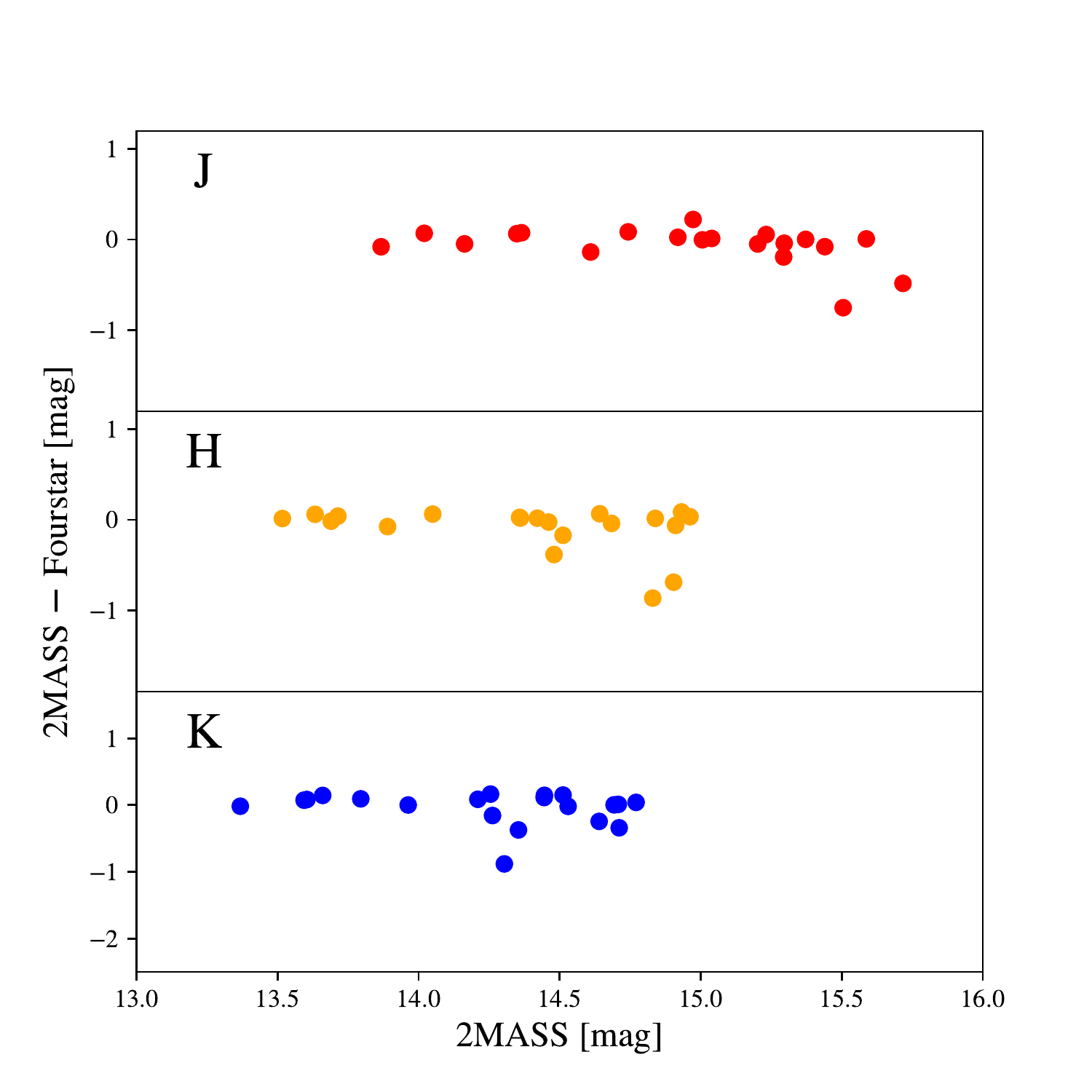"}{(b)}}
\caption{Figure (a) shows the magnitude offsets between our IMACS instrumental photometry and \textit{HST} calibrated photometry. Previous studied using the same \textit{HST} data showed the $TRGB_{F814W}$ to be around 20.91~mag \citep{2019MNRAS.490.5538A}, so we restricted our search for stars in both the IMACS $I$-band data and the \textit{HST} archival data to be between $20.80<m_{F814W}<21.30$ mag in order to most accurately measure the magnitudes of our targeted TRGB stars. The average magnitude offset was determined to be $<F814W-I>=8.350 \pm.003$ mag. Figure (b) shows the magnitude offsets between our FourStar photometry and 2MASS standards. 
\label{fig:f2}}
\end{figure}

The last step of the calibration process was to bring the IMACS instrumental magnitudes onto the Vegamag magnitude system. This was done by finding stars in both the calibrated \textit{HST} data and the IMACS instrumental data and computing the mean magnitude offset between them, and then applying that offset to the IMACS data. In order to prioritize the accuracy of the TRGB stars' magnitudes, stars were chosen based on their $I$-band magnitude near the TRGB. 
We chose a range of magnitudes based off of the reported WLM TRGB value of $m_{F814W}=20.91$ mag from \cite{2019MNRAS.490.5538A}, a study that used the same \textit{HST} data as this study. Thus, we adopted a range of 20.80 and 21.30~mag for our calibration stars. In total, 23 overlapping TRGB stars were found in both the \textit{HST} ACS/WFC footprint and the IMACS data. We show all of the stars' calculated magnitude offsets in Figure \ref{fig:f2}. The mean magnitude offset between the \textit{HST} calibrated magnitudes and IMACS instrumental magnitudes was determined to be $8.350 \pm0.003$ mag. This offset was applied to the instrumental magnitudes from IMACS.

The photometry was also cleaned of extended objects through the use of photometric quality cuts on the following \textsc{daophot} parameters: the photometric uncertainty $\sigma$, the sharp-parameter ($sharp$), and the chi-parameter $\chi$. Cuts were made based on a constant+exponential function as a function of magnitude. The specific parameters for the exponential functions can be found in the Appendix \ref{App:qualitycut}. 

\begin{figure*}
\centering
\includegraphics[width=0.85\textwidth]{"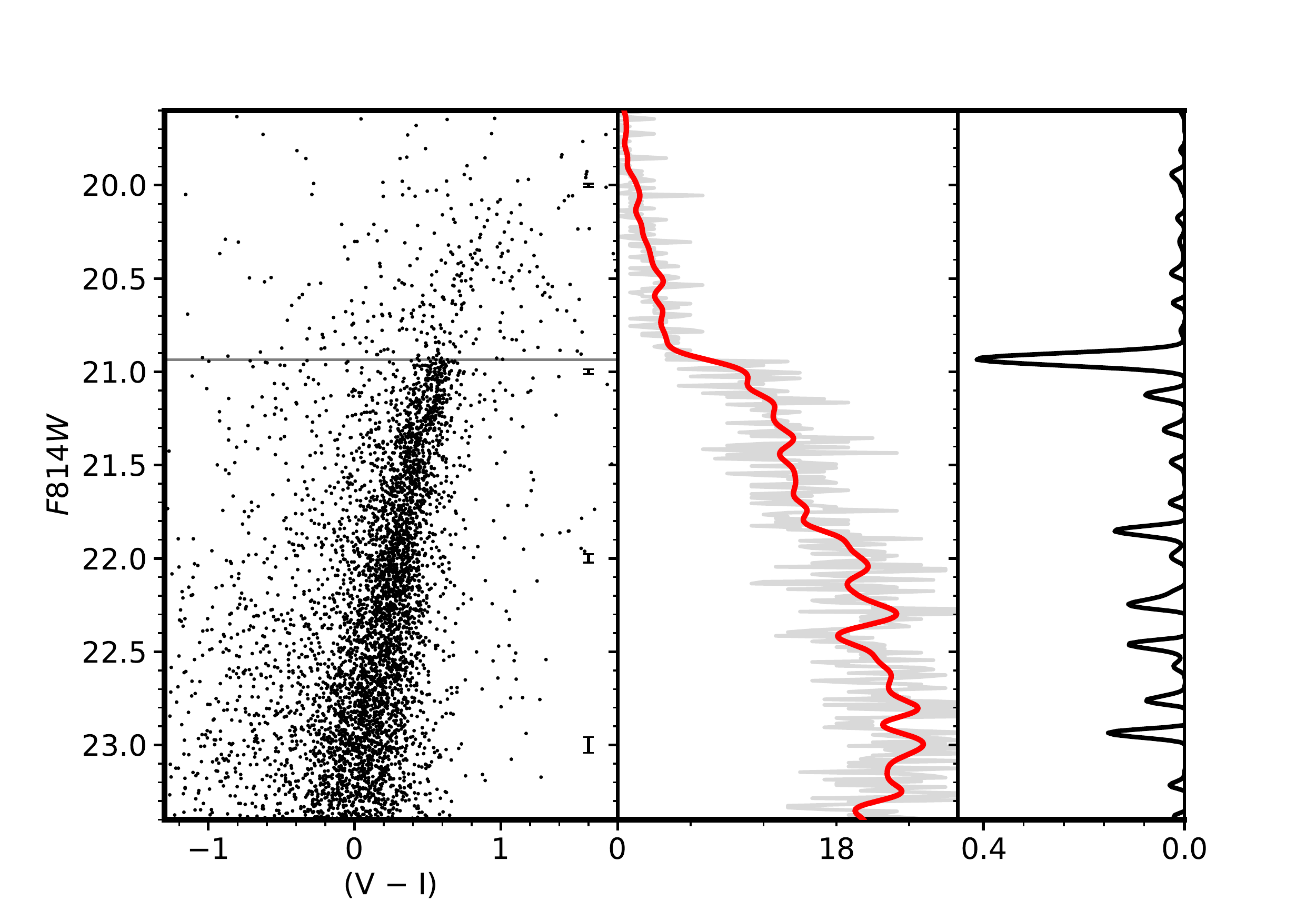"}
\caption{Color Magnitude Diagram for the optical $I$ band. The TRGB detection is clear and distinct. Error bars show uncertainties at 1 mag intervals. The middle panel shows the $I$-band GLOESS-smoothed luminosity function, and the right panel shows the edge response function.
\label{fig:f3}}
\end{figure*}

After making photometric quality cuts, we plotted a calibrated $I$ vs. $(V-I)$ color magnitude diagram. In Figure \ref{fig:f3}, we show this final color magnitude diagram, $I$-band luminosity function, and TRGB edge detection. The plotted stars are all in the halo of WLM, eliminating crowding effects and AGB contamination from the disk. 
In Section \ref{sec:trgb}, we describe our methods for the TRGB detection process for the optical $I$ band, and our subsequent tip detections in the near-infrared.

\subsection{JHK Photometry}\label{subsec:jhk}

\subsubsection{Magellan-Baade Telescope: FourStar} \label{subsubsec:mbtfourstar}
Ground-based $JHK$ data were taken at the $6.5$m Magellan-Baade telescope at Las Campanas Observatory with the FourStar Infrared Camera \citep{2013PASP..125..654P}. The FourStar imager has a field of view of $10.8\arcmin \times 10.8\arcmin$ and a resolution of 0.159~arcsec/pixel. We took three separate sets of observations, detailed in Table \ref{tab:observations}. The first set of observations, taken September 7, 2011, targeted the near-infrared TRGB and JAGB stars. The second set of observations, taken on October 5, 2011, also targeted JAGB stars. Finally, the third set of observations, used for calibration purposes, was taken November 13, 2019. 

\subsubsection{JHK Photometry}\label{subsubsec:jhkphot}
To obtain PSF photometry on the $JHK$ data, \textsc{daophot} was run on each frame using the standard procedure outlined in Section \ref{subsec:phot}. The next step was then to calibrate our instrumental $JHK$ photometry onto the 2MASS filter system, adopted for the $TRGB_{JHK}$ calibrations by \cite{2018ApJ...858...12H}.  We utilized the 2MASS single-epoch photometry in the $JHK$ filters.

The infrared $JHK$ data taken September 7, 2011 and October 5, 2011 were first moved onto the same instrumental magnitude system as the shallower dataset taken November 13, 2019 as the original datasets were too deep and thus  contained only one or two unsaturated stars in common with the 2MASS data. These data were then calibrated using data from 2MASS by finding stars in common with the 2MASS data and the FourStar dataset in a similar fashion as that described in Section \ref{subsubsec:calibrationVI}. See Figure \ref{fig:f2} for a plot of the FourStar to 2MASS calibration. We applied photometric quality cuts in an identical fashion to our $VI$ data to clean it of extended objects. The specific exponential function forms used can be found in Appendix \ref{App:qualitycut}.

In Section \ref{subsec:trgbIR} and Section \ref{sec:jagb}, we describe how we use our $JHK$ data for our near-infrared TRGB detections and JAGB measurement, respectively.

\subsection{Galactic Foreground Extinction and Internal Reddening Corrections}
WLM is at high galactic latitude ($b=-73.6^\circ$) and is therefore expected to have a low line-of-sight extinction. We determined the foreground extinction correction using the online IRSA Galactic Dust Reddening and Extinction tool\footnote{\url{https:/irsa.ipac.caltech.edu/applications/DUST/}}, which queries the \cite{1998ApJ...500..525S} full-sky Milky Way extinction maps and gives reddening estimates based on a location and rescaling from \cite{2011ApJ...737..103S}. The foreground extinction corrections for this study in the $I, J, H, K$ bands respectively amount to  $A_I=0.057, A_J=0.027, A_H=0.017$ and $A_K=0.012$~mag, assuming the \cite{1989ApJ...345..245C} reddening law, $R_v=3.1$. As per \cite{2019ApJ...885..141B}, we choose to adopt a systematic uncertainty on the extinction correction equal to half its value.

To measure possible internal reddening differences resulting from {\it in situ} dust of WLM, we split our field into three non-overlapping subsections. For each field, the luminosity function was first smoothed using a smoothing parameter of $\sigma_s=0.10$, which is slightly larger than that which was used for the entire field, $\sigma_s=0.075$ (discussed in Section \ref{subsec:trgbI}) in order to account for the smaller numbers of stars. We then calculated the $TRGB_I$ for each section. We found a variation of 0.03~mag between the three fields. Because the variations between fields is so small, we conclude that there is a negligible differential internal reddening effect, as this variation is most likely just due to small number statistics \citep[e.g., see][]{2017ApJ...845..146H}. 

\section{Tip of the Red Giant Branch} \label{sec:trgb}
The use of the TRGB as a distance indicator can be understood within the context of stellar evolution for low-mass red giant stars \citep{1997MNRAS.289..406S}. The beginning of helium burning by the triple-$\alpha$ process in the star's core is marked by a sharp discontinuity in the red giant branch luminosity function (at the location of the helium flash) \citep{1993ApJ...417..553L}. 

In the $I$ band, the luminosity of the TRGB is remarkably insensitive to color, and can used as a standard candle (without additional color corrections) \citep{2019ApJ...882...34F,2020arXiv200804181J}. The TRGB distance indicator method is also well understood theoretically \citep{1997MNRAS.289..406S} and is simple and empirically-based. Recent studies have also shown the TRGB can also be be used as a distance indicator in the near-infrared $JHK$ bands even though the TRGB becomes upward-sloping with increasing color \citep{2018ApJ...858...11M, 2018ApJ...858...12H}.

For all of our TRGB detections, we made sure to only include stars in the halo of WLM in order to minimize contamination of the TRGB by brighter intermediate-age AGB stars, {\it in situ} reddening from gas and dust in the disk, and to avoid crowding/blending in our photometry.

We first constructed four TRGB color magnitude diagrams: $J$ vs. $(J-K)$, $K$ vs. $(J-K)$, $H$ vs. $(J-K)$, and $I$ vs. $(V-I)$. As mentioned above, in the optical $I$ band, the TRGB has a nearly constant magnitude, with little to no dependence on age or metallicity  in the color range found for the RGB in WLM, whereas in the near-infrared, the TRGB is known to be upward sloping with increasing color. In Section \ref{subsec:trgbI}, we describe our measurement of the $I$-band TRGB, and in Section \ref{subsec:trgbIR}, we describe our subsequent measurements of the near-infrared TRGB zero point.

\subsection{Optical $I$-band TRGB} \label{subsec:trgbI}

\begin{deluxetable*}{ccll}
\tablenum{2}
\tablecaption{Error Budget for WLM \label{tab:errors}}
\tablewidth{1pt}
\tablehead{
\colhead{Source of Uncertainty} & 
\colhead{Value [mag] } & 
\colhead{$\sigma_{stat}$ [mag]} &
\colhead{$\sigma_{sys}$ [mag]} 
}
\startdata
Edge Detection & $~...$& 0.01 & 0.01 \\
STScI ACS F814W ZP & 25.517 & $~...$ & 0.02 \\
STScI ACS F814W Infinite Aperture Correction & 0.098 & $~...$ & 0.02 \\
Differential Aperture Correction & -0.107 & 0.009 & $~...$ \\
$<$ACS $-$ IMACS$>$ & 8.350 & 0.003 & $~...$\\
$F814W(TRGB)$ & 20.93 & 0.01 & 0.03\\
\hline\hline
$A_{F814W}$ & 0.057 & $~...$ & 0.029\\
$M^{TRGB}_{F814W}$ \citep{2020ApJ...891...57F} & -4.054 &  0.022 & 0.039\\
$\mu_0 (TRGB)_{F814W}$ & 24.93 & 0.02 & 0.06\\
\hline\hline
$J(TRGB)$ &19.85  &0.036& $0.017$ \\ 
$A_{J}$ & 0.027 & $~...$ & 0.014\\
$M^{TRGB}_{J}$ \citep{2018ApJ...858...12H} & -5.14 &  0.01 & 0.06\\
$\mu_0 (TRGB)_J$ & 24.96 & 0.04 & 0.06\\
\hline
$H(TRGB)$  &19.05 &0.038 & 0.022 \\
$A_{H}$ & 0.017 & $~...$ & 0.029\\
$M^{TRGB}_{H}$ \citep{2018ApJ...858...12H} & -5.94 &  0.01 & 0.06\\
$\mu_0 (TRGB)_H$ & 24.97 & 0.04 & 0.07\\
\hline
$K(TRGB)$ & 18.85& 0.036 & 0.030 \\
$A_{K}$ & 0.012 & $~...$ & 0.029\\
$M^{TRGB}_{K}$ \citep{2018ApJ...858...12H} & -6.14 &  0.01 & 0.06\\
$\mu_0 (TRGB)_K$ & 24.98 & 0.04 & 0.08\\
\hline\hline
$J(JAGB)$& 18.80 & 0.02 & 0.007\\
$A_J$ & 0.027 & $~...$ & 0.014\\
$M_J^{JAGB}$  \citep{2020arXiv200510792M}& -6.20 &0.01 & 0.04\\
$\mu_0 (JAGB)_J$ & 24.97 & 0.02 & 0.04\\
\enddata
\end{deluxetable*}

To detect the $I$-band TRGB, we followed the CCHP procedure standardized in \cite{2017ApJ...845..146H}. First, the $I$-band luminosity function was binned at the 0.01 mag level, and then smoothed using a non-truncated Gaussian kernel smoothing algorithm called GLOESS (Gaussian-windowed, Locally-weighted Scatterplot Smoothing) \citep{2004AJ....128.2239P,2017AJ....153...96M}  in order to reduce Poisson noise peaks (see the middle panel of Figure \ref{fig:f3}). We chose a smoothing parameter of $\sigma_s=0.075$ for the luminosity function. Then the smoothed luminosity function was convolved with a Sobel kernel [-1,0,+1] (first derivative) edge detection filter, producing a gradient approximation which has also been weighted by signal-to-noise. The resulting edge response shows where the slope of the luminosity function is the steepest, or where the first derivative of the luminosity function is the largest (see the right-most panel of Figure \ref{fig:f3}). The maximum of this edge response function marks the TRGB. 

Figure \ref{fig:f3} demonstrates our extremely clear tip detection in the $I$ versus $(V-I)$ data (distinct enough to be visually apparent to a few hundredths of a magnitude). We base our systematic and statistical error associated with our TRGB measurement on the artificial stars experiments performed in a previous CCHP study, \citep{2017ApJ...845..146H}. From their analyses, we have adopted a conservative estimate of the systematic error on our measurement of the TRGB,  $\sigma_{sys} = 0.01$~mag, and a conservative estimate of $\sigma_{stat}=0.01$~mag based on the width of our narrow edge-response function.  After including errors from the calibration procedure, we present a TRGB measurement in the optical of $m_{F814W}^{TRGB}=20.93$ $\pm0.01$ (stat) $\pm0.03$ (sys)~mag. Adopting the extinction correction of $A_{I}=0.057$ $\pm 0.02$ (sys)~mag and an absolute magnitude of $M_{F814W}=-4.054$ $\pm0.022$ (stat) $\pm0.039$ (sys)~mag from \cite{2020ApJ...891...57F}, we determine the distance modulus based on the $I$-band TRGB to be $\mu_0 (TRGB_{F814W}) =24.93$ $\pm 0.02$ (stat) $\pm0.06$ (sys)~mag.

\subsection{Near-Infrared TRGB}\label{subsec:trgbIR}

\begin{figure*}
\gridline{\fig{"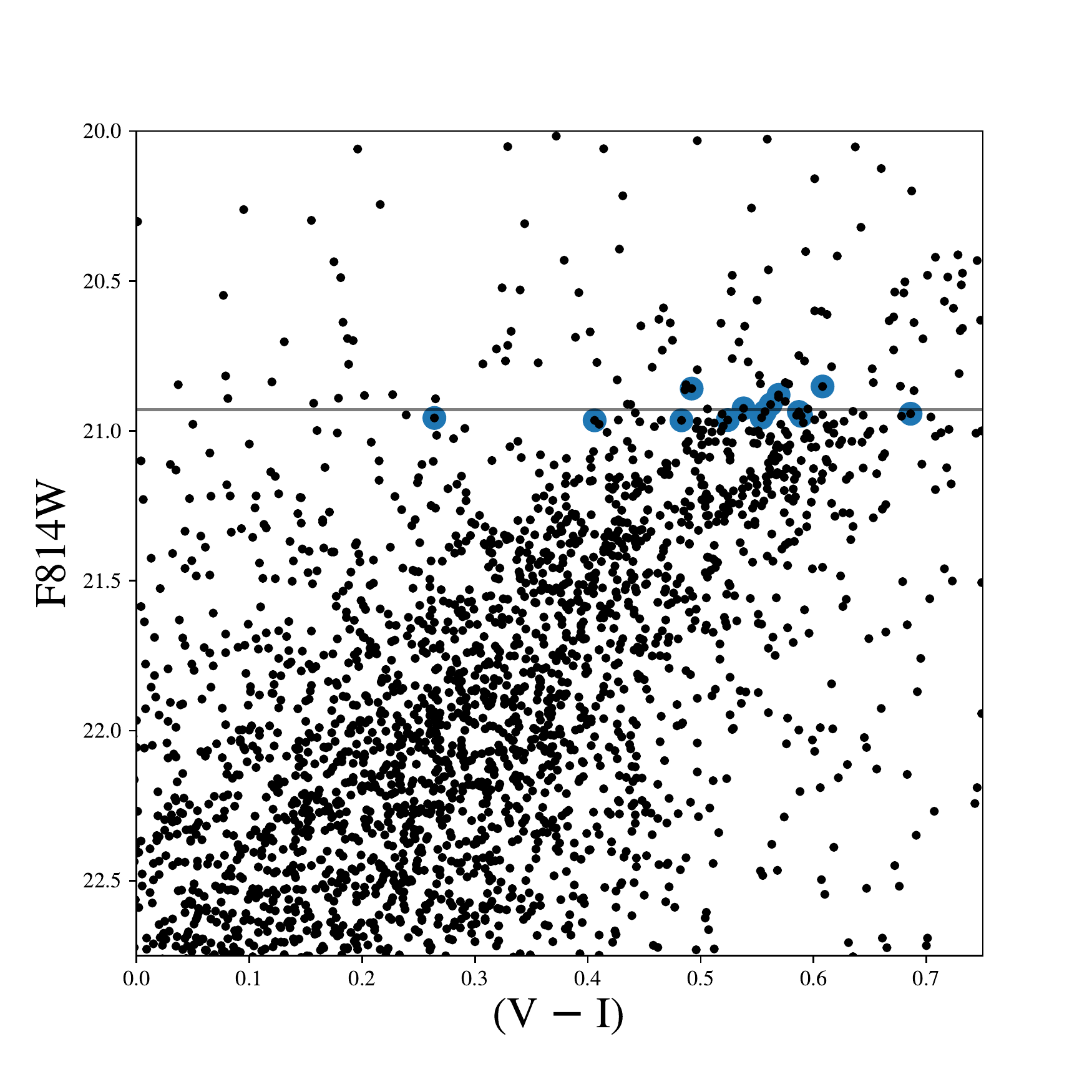"}{0.5\textwidth}{}
          \fig{"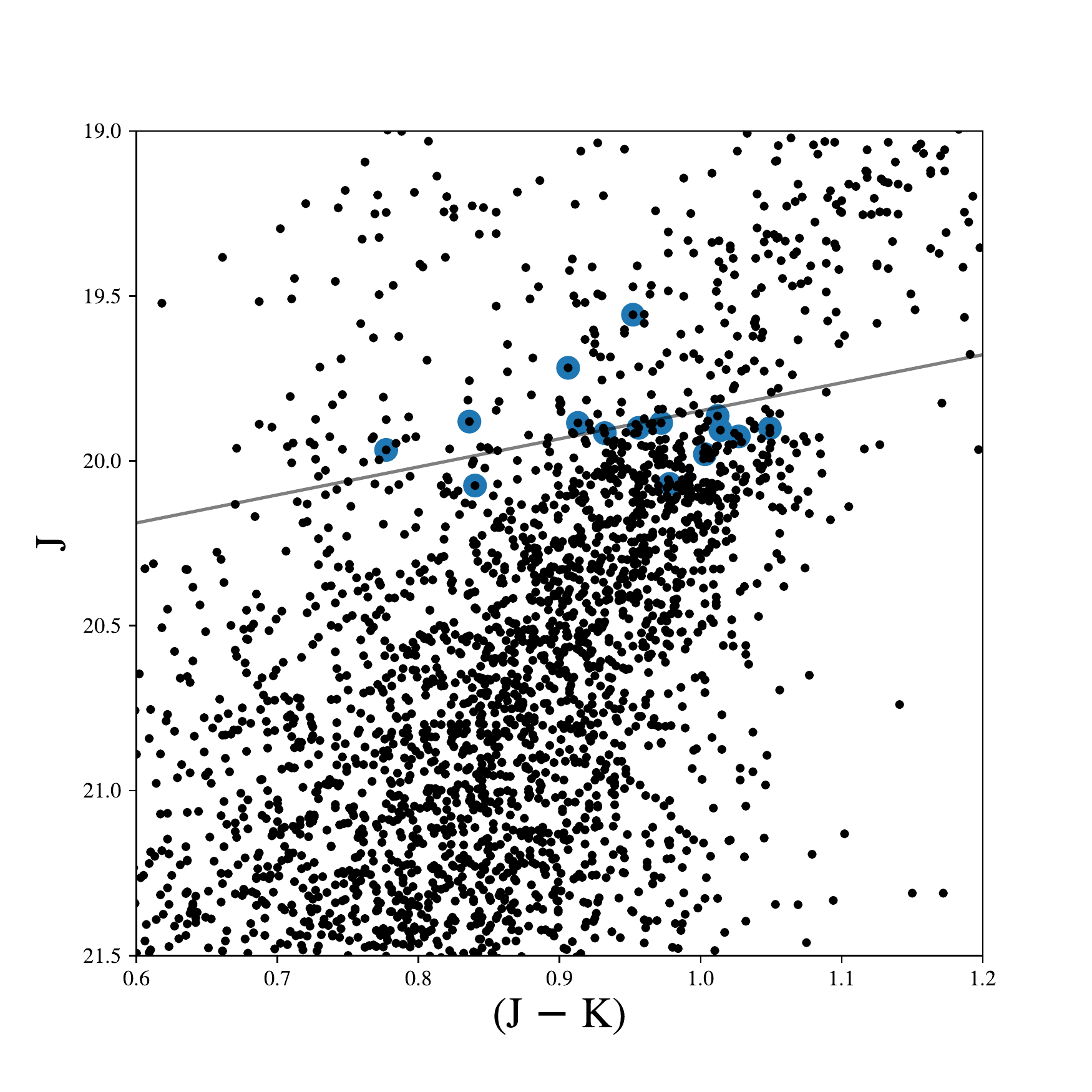"}{0.5\textwidth}{}}
\gridline{\fig{"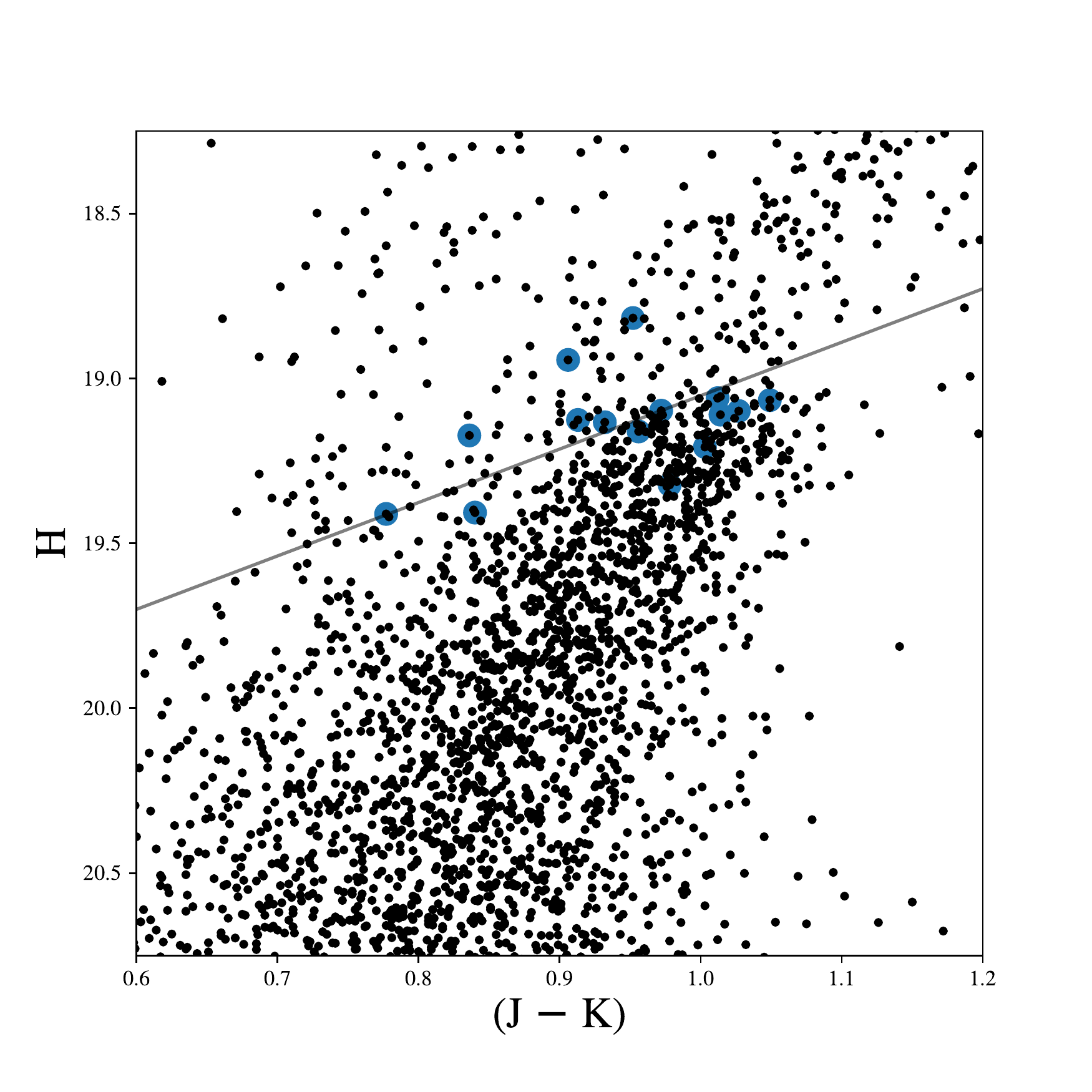"}{0.5\textwidth}{}
          \fig{"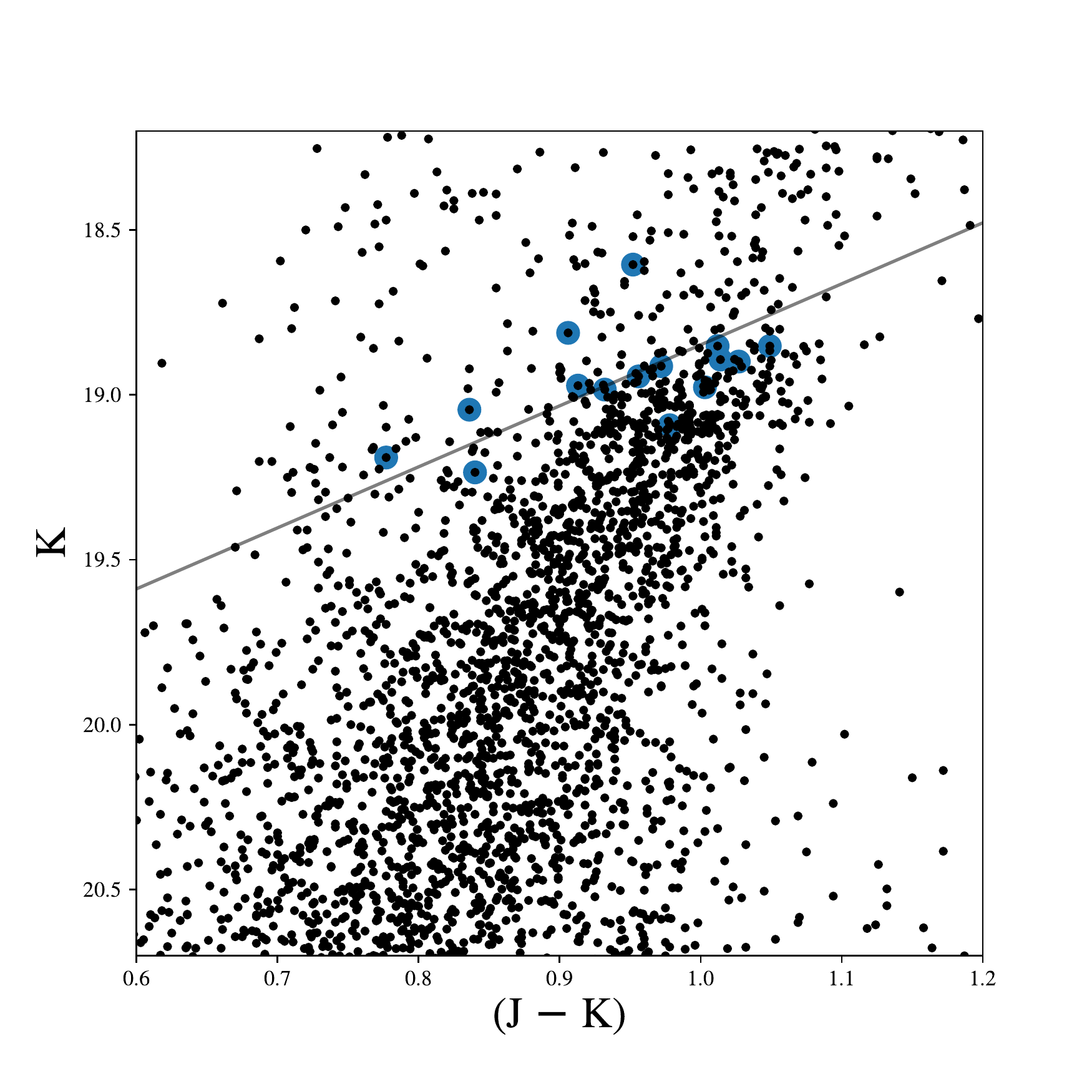"}{0.5\textwidth}{}}
\caption{Color Magnitude Diagrams for WLM in the Optical and Near-Infrared. We cross-identified 15 stars (blue points) near the TRGB in the $I$ band where our tip detection was the clearest, and then mapped them into $JHK$. We then fit zeropoints using pre-determined slopes by minimizing the scatter for $J$ vs. $(J-K)$, $H$ vs. $(J-K)$, $K$ vs. $(J-K)$ 
\label{fig:f4}}
\end{figure*}

Since the conception of the TRGB method with the use of a Sobel filter as described in \cite{1993ApJ...417..553L}, the majority of subsequent studies have used the optical $I$ band in their endeavors as the $I$-band TRGB is `flat' in magnitude and is largely independent in metallicity and color. On the other hand, whereas the $I$-band TRGB only has a unique zero point because its color dependence is 0, the near-infrared (NIR) TRGB is upward-sloping with color, and therefore has both a zero-point and slope term. 

The NIR TRGB is a relatively recently developed distance indicator. Only a few studies to date have employed the use of the NIR TRGB to measure distances to galaxies. \cite{2012ApJS..198....6D} first published NIR CMDs, indicating a promising future for the NIR TRGB as a distance indicator. \cite{2014AJ....148....7W} subsequently undertook a similar analysis and came to the same conclusion. \cite{2011AJ....141..194G} derived NIR TRGB distances to five galaxies in the local group. \cite{2018ApJ...858...12H} derived a NIR absolute calibration for the NIR TRGB for the 2MASS system, and \cite{2018ApJ...858...11M} subsequently used their calibration to publish a distance to IC~1613 in a companion paper. Finally, \cite{2020ApJ...898...57D} recently developed a new method for determining the NIR TRGB by fitting an n-dimensional Gaussian to photometry of TRGB stars.

Although use of the NIR TRGB is relatively new, there are two distinct advantages of employing the NIR TRGB over the optical. The first is that the TRGB is about 2 magnitudes brighter in the NIR than the optical, and the second is that line-of-sight extinction effects decrease for redder wavelengths.

To detect the NIR TRGB, we follow the procedure in \cite{2020ApJ...891...57F}: we first selected a set of ``tracer tip stars" in the $I$ band, as it is flat and our detection is well-defined. We then mapped those stars into the $JHK$ bands to obtain NIR CMDs. Finally, we fit a TRGB to the tracer stars in the NIR using predetermined slopes from \citep{2018ApJ...858...12H} to determine the TRGB zeropoint by minimizing the scatter for the zeropoint fit. This method is effective because the same stars defining the tip in the flat $I$ band are still the same stars that define the TRGB at other wavelengths.

We selected 15 tracer stars with an $I$-band mean equal to that of the peak magnitude of the $TRGB_I$. These stars were then identified in the $JHK$ data by using  \textsc{topcat} to match the catalogs by their $\alpha/\delta$. Next, these stars were plotted in color-magnitude diagrams for $J, H$  and $K$ vs $(J-K)$. Finally, a line was fit to the points for each CMD using the predetermined slopes from the absolute calibration of \cite{2018ApJ...858...12H} using a least-squares regression to find the zeropoint. We present the NIR TRGB fits overlaid on their respective CMDs in Figure \ref{fig:f4}, along with the stars used for mapping in the $I$ band. We found:
\begin{equation}
m_J^{TRGB}=19.85-0.85[(J-K)_o - 1.00]
\end{equation}
\begin{equation}
m_H^{TRGB}=19.05-1.62[(J-K)_o - 1.00]
\end{equation}
\begin{equation}
m_K^{TRGB}=18.85-1.85[(J-K)_o - 1.00]
\end{equation}
Our fitted zeropoints had errors on the mean of 0.036, 0.038, and 0.036 mag for the fits on $JH$ and $K$, respectively.

We determined distance moduli based on absolute calibrations from \cite{2018ApJ...858...12H}, which we repeat here:
\begin{equation}
M_J^{TRGB}=-5.14-0.85[(J-K)_o - 1.00]
\end{equation}
\begin{equation}
M_H^{TRGB}=-5.94-1.62[(J-K)_o - 1.00]
\end{equation}
\begin{equation}
M_K^{TRGB}=-6.14-1.85[(J-K)_o - 1.00]
\end{equation}
The quoted errors on the zeropoints are $\pm0.01$ mag (stat) and $\pm0.06$ mag (sys). 

We then applied extinction corrections, and factored in the average photometric error for each of the filters near the TRGB: 0.017, 0.022, and 0.030 mag for the $JH$ and $K$ filters respectively. Our final NIR TRGB distance moduli are: $\mu_0 (TRGB_J)=24.96 \pm 0.04$ (stat) $\pm0.06$ (sys) mag, $\mu_0 (TRGB_H)=24.97 \pm 0.04$ (stat) $\pm0.07$ (sys) mag, $\mu_0 (TRGB_K)=24.98 \pm 0.04$ (stat) $\pm0.08$ (sys) mag. Averaging these values gives a NIR TRGB value of $\mu_0 (TRGB_{NIR})=24.98 \pm 0.04$ (stat) $\pm0.07$ (sys) mag.

\section{The JAGB Method} \label{sec:jagb}
The JAGB method is a precise and accurate distance indicator recently introduced in \cite{2020arXiv200510792M}, which followed on that of \cite{2000ApJ...542..804N, 2001ApJ...548..712W}. Recently, an independent study by  \cite{2020MNRAS.495.2858R} also indicated that the peak of the JAGB luminosity function had to the potential to be a powerful standard candle.
JAGB stars are thermally-pulsating AGB stars. During pulsations, the convective envelope penetrates deep into the star, bringing material to the surface. Each dredge-up phase penetrates increasingly deeper, and by the star's third-dredge-up phase (and later), the convective envelope brings carbon from the He-burning shell to the stellar surface \citep{2003agbs.conf.....H}.

Because of the carbon in JAGB stars' atmospheres, they are photometrically distinct from and redder than oxygen-rich AGBs in color-magnitude space, and in the NIR they have well-defined limits in color and magnitude.  To delineate JAGB stars in color, \cite{2020arXiv200510792M} chose conservative color cuts of $1.30 < (J-K) < 2.00$~mag,  following \cite{2000ApJ...542..804N}. Stars bluer than $(J-K) = 1.30$ mag are primarily oxygen-rich AGB stars, but also include carbon core-burning TP-AGBs that have not yet `dredged-up' enough carbon to become JAGB stars. Stars redder than $(J-K) = 2.00$ mag are extreme carbon stars with winds in which dust is being formed, leading to the star being increasingly obscured by its own circumstellar envelope \citep{2000ApJ...542..804N}. The third and later dredge-up phases of older and less-massive stars are not sufficiently deep to dredge up carbon; while the younger and more massive stars have such hot convective envelopes that the carbon is burned into nitrogen before it has time to reach the surface \citep{2007A&A...469..239M, 2003agbs.conf.....H}. From an empirical perspective, there exists an intermediate range of masses (and corresponding luminosities) for AGB stars that have carbon-enriched atmospheres that we see as JAGB stars.
As demonstrated in \cite{2020arXiv200510792M}, the above physical constraints manifest themselves in the NIR as JAGB stars, having a low-dispersion mean magnitude in the $J$ band, which is independent of color, and therefore qualifies as a standard candle. 

The JAGB method has the potential to be an important and widely applicable distance indicator. \cite{2001ApJ...548..712W} found that most stars within the JAGB color limits of $1.30<(J-K)<2.00$ are variable.
The variability  of JAGB stars increases the scatter in the JAGB luminosity function, but does not contribute a systematic error \citep{2020arXiv200510793F}. The statistical error can be lowered significantly by using a large number of JAGB stars. 
Empirically, the distribution of JAGB stars was found to have an intrinsic dispersion of only $\pm$0.20~mag \citep{2001ApJ...548..712W, 2020arXiv200510793F}. The statistical error on the mean magnitude then decreases as $\pm0.20/\sqrt{N_c}$ mag, where $N_c$ is the number of JAGB stars in the observed JAGB luminosity function.
Additionally, in the $J$ band, systematic errors due to metallicity and reddening effects appear to be small \citep{2020arXiv200510793F}. 

Furthermore, the JAGB method has several advantages over other local distance indicators, including the TRGB method and the Cepheid Leavitt law. First, in the near-infrared, JAGB stars are about one magnitude brighter than the TRGB in the J band, so they can be detected out to appreciably greater distances. Second, the JAGB method needs only one epoch of observations (like the TRGB), whereas the Cepheids require upward of a dozen observations, spread out over many months, in order to detect variability and measure periods, amplitudes, mean magnitudes and mean colors. Finally, JAGB stars are found in all types of galaxies, unlike Cepheids, which are only found in the star-forming disks of late-type spiral and irregular galaxies. Therefore, the JAGB method can be applied to a heterogeneous sample of type Ia supernova host galaxies, including lenticulars and ellipticals.

\subsection{Measurement of the JAGB Magnitude for WLM}
\begin{figure}
\centering
\includegraphics[width=\columnwidth]{"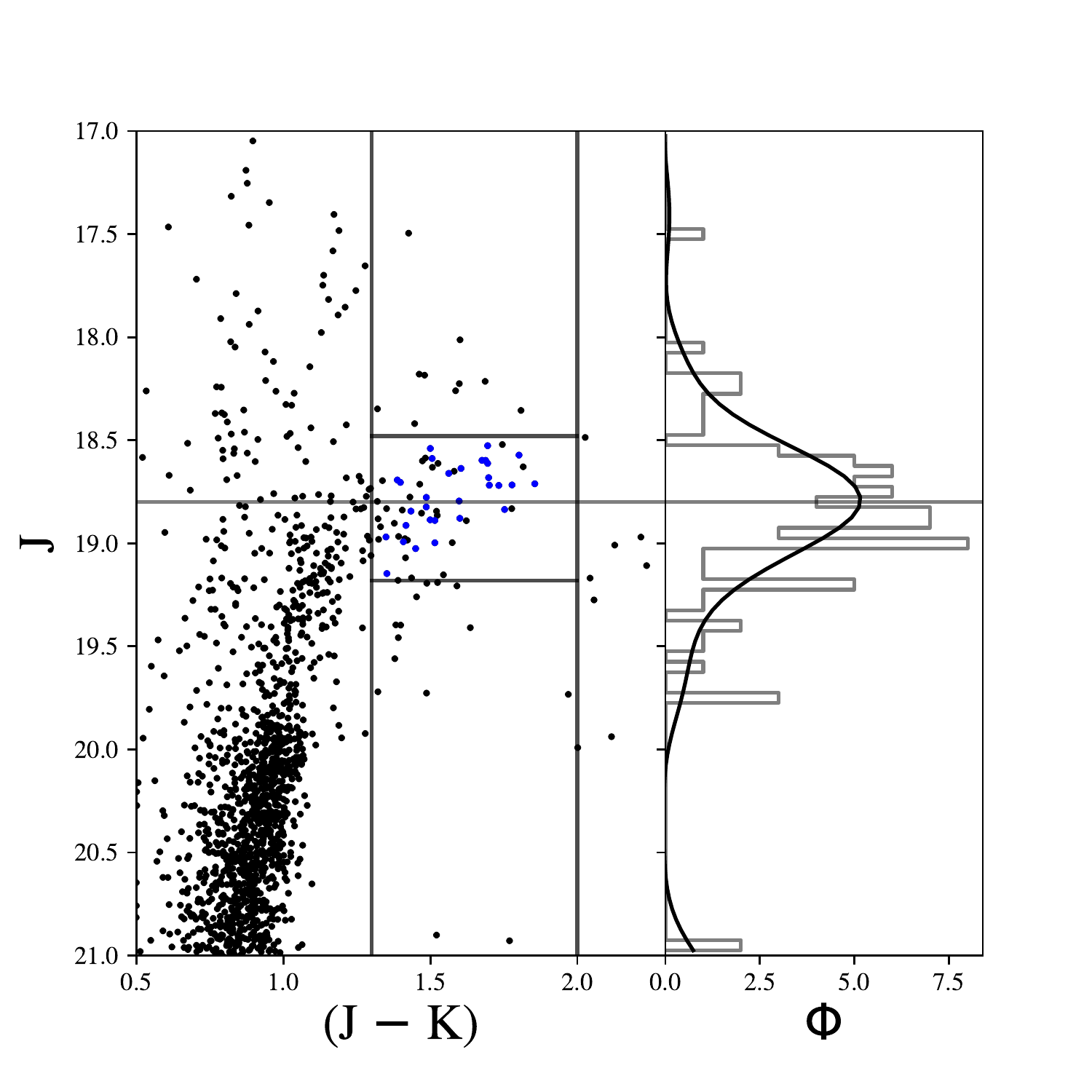"}
\caption{Color Magnitude Diagram (left) and the smoothed luminosity function from a bin size of 0.05 mag (right) of the JAGB star distribution. This data is based on two epochs of observing and are temporally averaged. The horizontal line represents the mean $J=18.80$~mag. Vertical lines show the color cuts of $1.30<(J-K)<2.00$ mag made on the LF in order to avoid oxygen-rich AGB stars on the blue side, and K branch stars on the red side. The two horizontal lines at $J=19.18$ and $J=18.48$ mag represent upper and lower magnitude limits for the JAGB stars distribution. The blue points represent the 30 stars time-averaged with the \cite{2007A&A...466..501V} near-infrared dataset of AGB/carbon stars in their study of WLM.
\label{fig:f5}}
\end{figure}

WLM was identified as a good candidate for the JAGB method because of its high number of carbon stars, quantified by a high carbon to M star ratio (C/M). 
Galaxies with high C/M ratios are also known to have low metallicities \citep{1986ApJ...305..634C}. WLM's low metallicity of [Fe/H]$ = -1.45 \pm0.2$~dex \citep{1997AJ....114..147M} made it a good candidate for measuring both a TRGB distance in the halo and a JAGB distance in the outer disk.

In Figure \ref{fig:f5}, we show the color-magnitude diagram of our $J$ vs $(J-K)$ data. The  two epochs of FourStar data, which were taken about a month apart (on September 7, 2011 and October 5, 2011), have been averaged together. Grey lines show the adopted color cut-offs used to select JAGB stars; i.e., $1.30 < (J-K) < 2.00$~mag. In Paper II, \cite{2020arXiv200510793F} measured distance moduli using the mean J-band magnitudes of the JAGB populations. We adopt that same measure here, while noting that future studies of galaxies with larger statistics will be needed to explore the possible effects of star formation histories and metallicities, for example, on the higher moments and multi-modal substructure in the overall JAGB luminosity function. In WLM we only have 59 JAGB stars selected from Figure 5, centered on a mean value of $J = $ 18.83 $\pm$ 0.35~mag, where the upper and lower limits are adopted from
Weinberg \& Nikolaev (2001) based on several thousand, single-epoch observations of JAGB stars in the LMC.

Finally, we note that \cite{2005A&A...434..657B} identified 77 carbon stars in WLM observations from \cite{2007A&A...466..501V}, which we cross-matched to our dataset. After applying our JAGB color and magnitude cuts to their data, we identified 30 stars in common between our two datasets. In Figure \ref{fig:f5}, we have plotted these stars in blue.  We have averaged our FourStar $J$ and $K$ magnitudes with the \cite{2007A&A...466..501V} data.

The mean apparent magnitude of the JAGB distribution is $m_J=18.80$ mag, with a dispersion of $\pm 0.17$ mag. Thus, the 59 JAGB stars contribute an error on the mean of $\pm0.17 /\sqrt{59}=0.02$ mag. The mean photometric error of stars in the $J$ band in the JAGB distribution yields an error of $\sigma_{sys}=0.007$ mag.  After applying an extinction correction of $A_J=0.027$ mag, and using the absolute magnitude calibration from \cite{2020arXiv200510792M} of $M_J = -6.20 \pm0.01$ (stat) $\pm0.04 $ (sys) mag, we determine a distance modulus of $\mu_0$ (JAGB) $=24.97 \pm0.02$ (stat) $\pm 0.04$ (sys) mag.

\section{Independent Distance Comparisons} \label{sec:comparison}

\begin{figure}
\centering
\includegraphics[width=\columnwidth]{"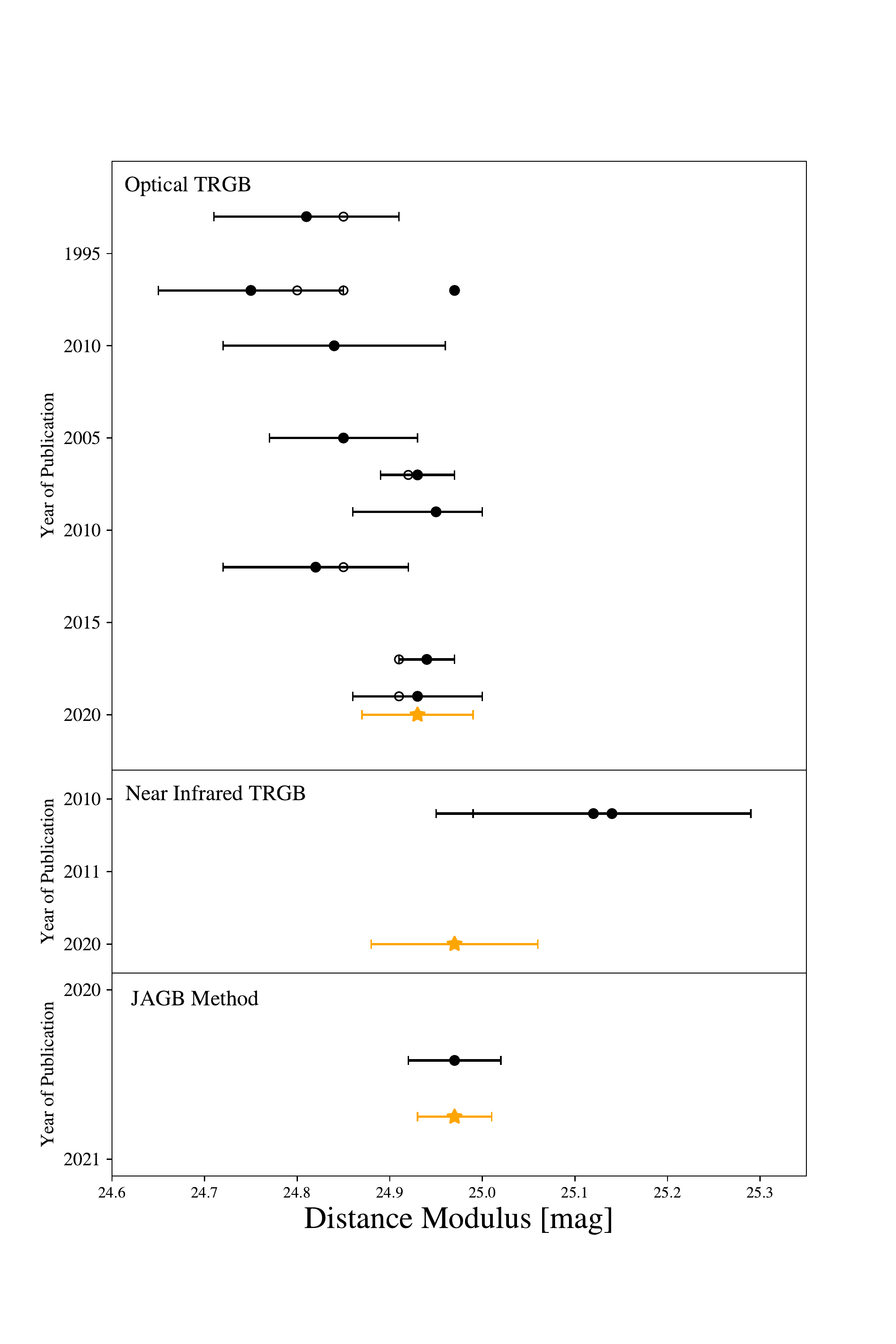"}
\caption{Previous distance modulus estimates to WLM. Filled circles represent as published values and open circles represent re-calculated standardized values using $M_{F814W}=-4.054$ \citep{2020ApJ...891...57F} and $A_I-0.057$ (NED). Papers from which these values were taken from are compiled in Table \ref{tab:history}. Gold stars represent values from this study.
\label{fig:f6}}
\end{figure}

In this section, we compare our JAGB and TRGB distances against previously published, independent measurements. A compilation of these distance moduli and their references can be found in Table \ref{tab:history}. A visual compilation can be found in Figure \ref{fig:f6}; filled circles represent original published values and their error bars, and open circles represent re-calculated values standardized to the \cite{2020ApJ...891...57F} TRGB calibration in the $I$ band and a foreground extinction value of $A_I=0.057$ mag. Finally, we present a Cepheid distance comparison in Section \ref{sec:cephcompare}.

\subsection{TRGB Distance Comparisons} \label{subsec:trgbcompare}
Since 1993, 11 TRGB distances have been measured in the $I$ band, one in the $J$ band, and one in the $K$ band for WLM. In Figure \ref{fig:f6}, we show our TRGB distance moduli compared with previous papers' measurements. The open dots represent re-calculated distance moduli normalized to the \cite{2020ApJ...891...57F} $I$-band calibration and a foreground extinction correction of $A_I=0.057$ mag (NED). The filled dots represent the original values and their errors quoted in the paper. 

We find the average value of the 11 published $TRGB_I$ distance moduli to be $\mu_0=24.87 \pm0.08$ mag. This is in agreement, at the 1-$\sigma$ level, with our measured value of $\mu_0 (TRGB_{F814W})=24.93 \pm0.02$ (stat) $\pm0.06$ (sys) mag.

For the NIR TRGB, \cite{2011AJ....141..194G} measured a near-infrared TRGB distance to WLM; their average value of $25.13 \pm 0.16$ mag differs by about 1-$\sigma$ from our measured value of $24.98 \pm 0.04$ (stat) $\pm0.08$ (sys) mag.

\subsection{JAGB Distance Comparison} \label{sec:jagbcompare}

Only one previous study has used the JAGB method to determine the distance to WLM \citep{2020arXiv200510793F}. They located known carbon stars found by \cite{2005A&A...434..657B} in $JK$ data obtained by \cite{2007A&A...466..501V} using ESO's New Technology Telescope. They found $J = 18.80 \pm0.25$ mag, with a true distance modulus of $\mu_0$ (JAGB) $=24.97 \pm0.05$ (stat) mag agreeing (likely fortuitously, exactly) with our value of $\mu_0$ (JAGB) $=24.97 \pm0.02$ (stat) $\pm 0.04$ (sys) mag.

\subsection{Cepheid Distance Comparison} \label{sec:cephcompare}

\begin{figure*}[p]
\centering
\includegraphics[width=\textwidth]{"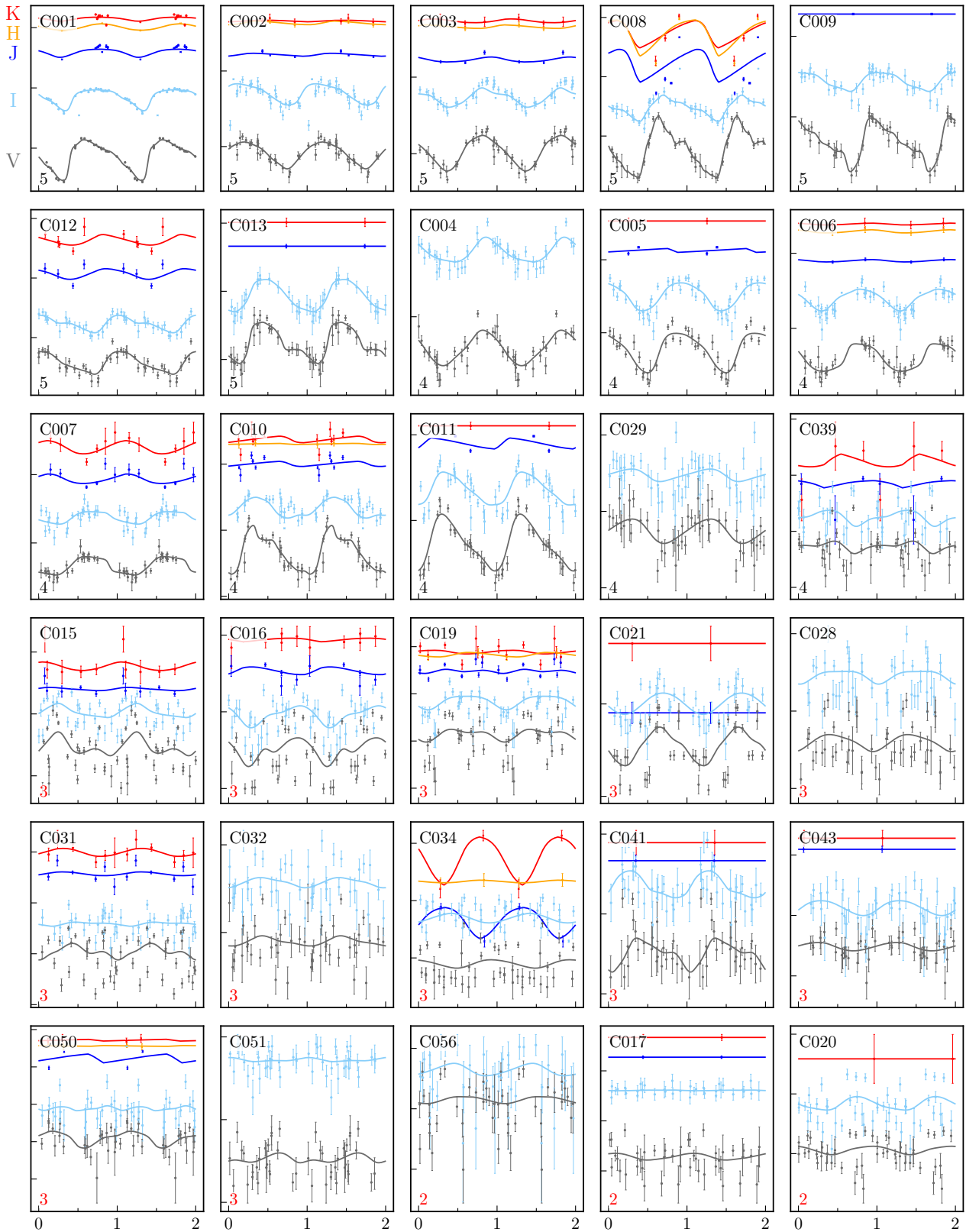"}
\caption{GLOESS-fit light curves in $VIJHK$ when available for the 60 Cepheids from \cite{2007AJ....134..594P}. Bands are distinguished by color, as labelled next to C001 at the top left. Cepheid numbers from the original publication are labeled in the upper-left corner of each subplot, and our quality rankings (where 5 = highest quality; 0 = lowest quality) are shown in the lower left. Red quality numbers indicate Cepheids excluded from the final sample. The light curve data shown in this figure are available as the Data behind the Figure.
\label{fig:f7}}
\end{figure*}

\begin{figure*}[p]
\figurenum{6. cont}
\centering
\includegraphics[width=\textwidth]{"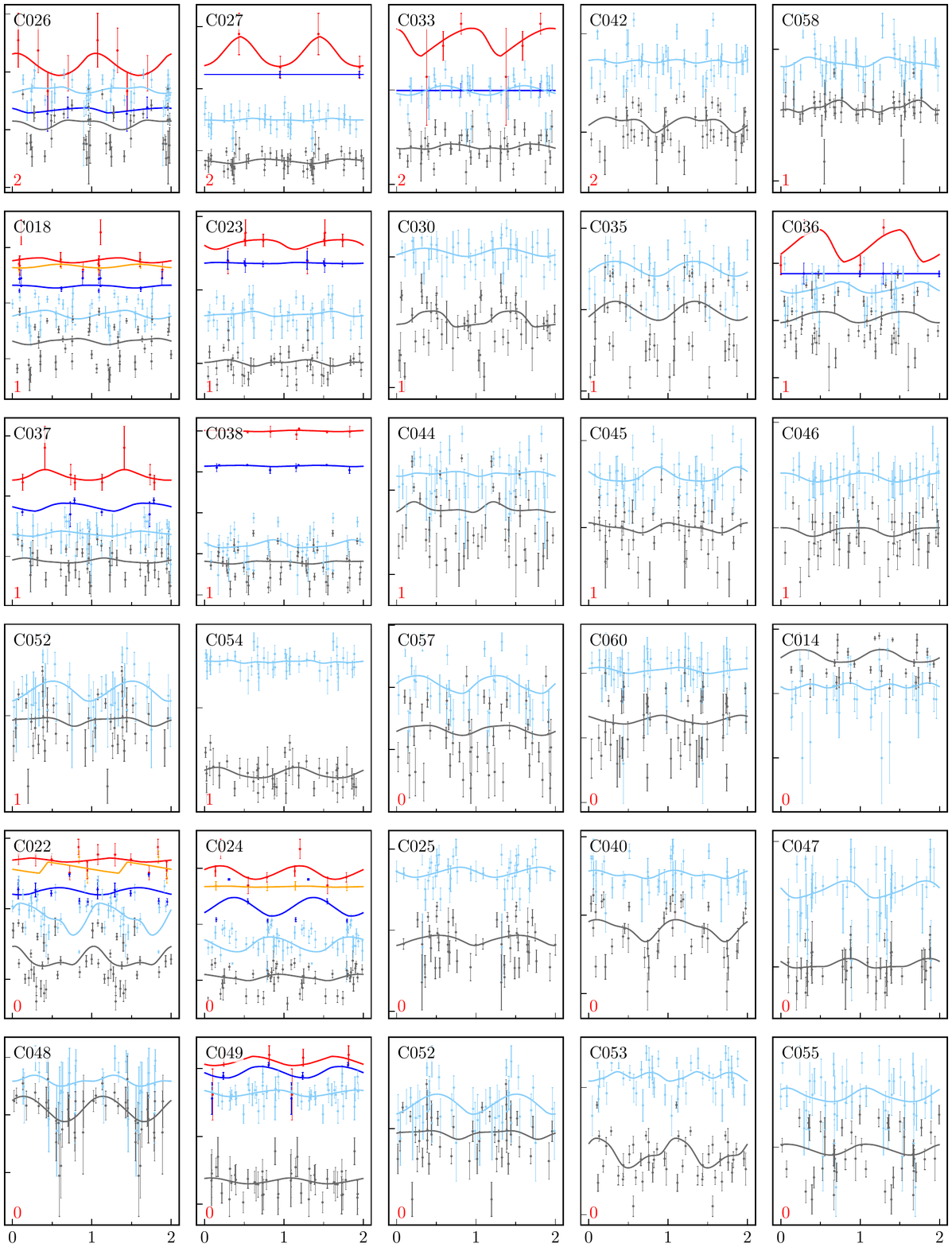"}
\caption{}
\end{figure*}

\cite{1985AJ.....90.1464S} (SC85) first reported the discovery of 15 short period Cepheids in WLM. These discoveries were based on 30 epochs of deep photographic $B$ band observations from the Palomar 5m and duPont 2.5m telescopes. Despite substantial effort spanning over 30 years of observation, their longest-discovered period was 9.6 days, and many Cepheids were close to the $B = 22.8$ mag detection limit of their photographic plates. It was another two decades before WLM was subject to a variable star search using modern detectors. \cite{2007AJ....134..594P} (P07) commenced a follow up search in the $V$ and $I$ bands using the Warsaw 1.3m telescope, and \cite{2008ApJ...683..611G} (G08) followed up in the $J$ and $K$ bands using infrared cameras on the New Technology and Baade Magellan telescopes. P07 found 60 Cepheids, recovering all but 2 of the original 15 discovered in SC85. Notably, they reported the first long-period Cepheid in WLM. SC85 originally reported Cepheid V12 with a period of 7.94 days, but P07 revised this to a significantly longer period of 54.7 days. G08 reported sparsely sampled (typically 3) $J$ and $K$ observations for 24 of the Cepheids in the P07 sample. We were able to add 6 new $JHK$ observations to each of 12 Cepheids taken on two nights in 2011 and 2019, as discussed above in Section \ref{subsubsec:mbtfourstar}. Additionally, \cite{2007A&A...466..501V} reported single phase points in the $J$ band for 5 Cepheids.

Merging these data sets and adopting the periods determined by P07, we GLOESS fit the light curves. We automated the selection of the GLOESS smoothing parameter $\sigma_{s}$, assigning it based on the number of points and the average error of those points in each band. These initial fits were visually inspected, and bands with $\sigma_{s}$ that significantly under-fit the data were refined by hand.

\begin{figure}
\centering
\includegraphics[width=\columnwidth]{"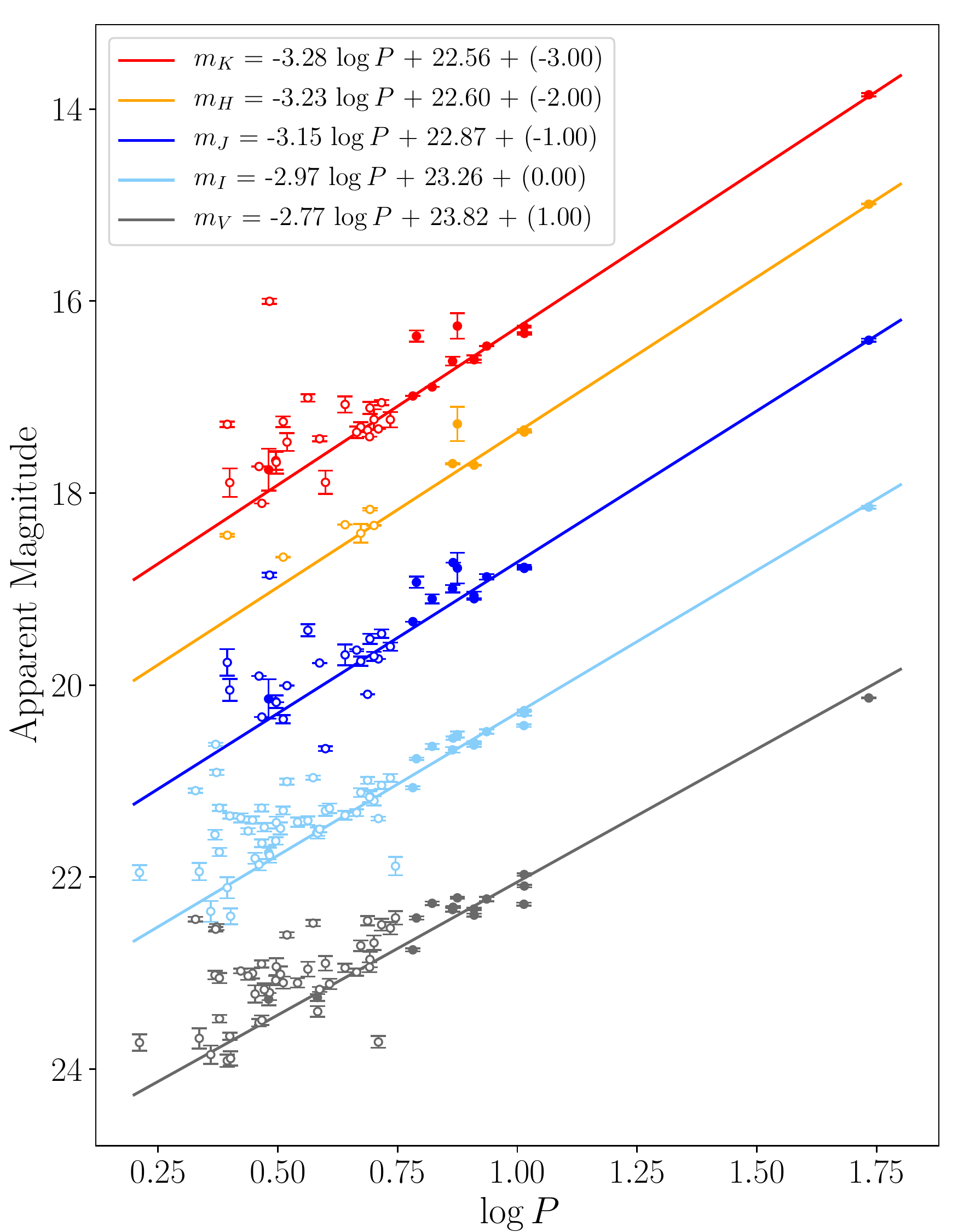"}
\caption{Period Luminosity relations in $VIJHK$ for WLM Cepheids with quality rankings of 4 or 5. The legend displays fits to the fixed slopes of \cite{2012ApJ...759..146M} in each band. Open circles indicate excluded data. It is clear that our quality cut closely resembles a period cut at $\log P \sim 0.8$.
\label{fig:f8}}
\end{figure}

Each of the Cepheids was ranked on a quality scale of 0 to 5 according to the amount of phase coverage, overall data quality, and the GLOESS goodness-of-fit across the bands. No period cuts were applied, but the quality rankings tracked period and apparent magnitude fairly closely, as might be expected. All 60 Cepheids are shown with our GLOESS fits and quality rankings in Figure \ref{fig:f7}.

From the final GLOESS fits, we calculated intensity-averaged mean magnitudes in all five bands. However, we should note that the archival $V$ band data from P07 was discovered to be incomplete when compared to the published light curve plots. Due to a server failure, the rest of the full data set was unavailable from the authors when contacted. We elected to still use the incomplete $V$ data since it is sufficient for calculating mean magnitudes, and we obtain similar values to the original publication.

We constructed five sets of multi-wavelength PL relations by performing progressive quality cuts. The largest set included all Cepheids, and the smallest data set contained only the 14 best Cepheids (ranked 4 and 5). We used this highest-quality subsample for determining our final distance modulus relations.

We fit the fixed slopes of \cite{2012ApJ...759..146M} to our Cepheid subsample and used their Milky Way intercepts to determine wavelength-dependent apparent distance moduli, plotted in Figure \ref{fig:f8}. Moduli and corresponding errors for all quality cutoffs are given in Table \ref{tab:DMs}. The \cite{1989ApJ...345..245C} extinction curve was fit to the distance moduli using a chi-squared grid minimizer, allowing the true distance modulus and reddening to vary. 

\begin{deluxetable*}{cccccccr}
\tablenum{3}
\tablecaption{Cepheid Distance Moduli Dependence on Quality Cutoffs\label{tab:DMs}} \tablewidth{1pt}
\tablehead{
	\colhead{Quality cutoff} & 
	\colhead{$\mu_{V}$} &
	\colhead{$\mu_{I}$} &
	\colhead{$\mu_{J}$} &
	\colhead{$\mu_{H}$} &
	\colhead{$\mu_{K}$} &
	\colhead{$\mu_{0}$}  & 
	\colhead{$E(B-V)$}
}
\startdata
all data & $25.03 \pm0.08$ & $24.90 \pm0.08$ & $24.97 \pm0.06$ & $24.82 \pm0.05$ & $24.93 \pm0.05$ & $24.96\pm0.05$ & $-0.02 \pm 0.07$ \\
$\geq 1$ & $25.03 \pm0.09$ & $24.91 \pm0.09$ & $24.97 \pm0.07$ & $24.85 \pm0.05$ & $24.97 \pm0.05$ & $24.96\pm0.05$ & $-0.01 \pm 0.07$ \\
$\geq 2$ & $25.03 \pm0.07$ & $24.89 \pm0.10$ & $24.96 \pm0.06$ & $24.94 \pm0.05$ & $25.04 \pm0.06$ & $24.94 \pm 0.04$ & $0.03 \pm 0.06$ \\
$\geq 3$ & $25.00 \pm0.08$ & $24.89 \pm0.10$ & $24.97 \pm0.06$ & $25.00 \pm0.04$ & $25.10 \pm0.04$ & $24.92 \pm 0.03$ & $0.06 \pm 0.05$ \\
$\boldsymbol{\geq 4}$ & $\boldsymbol{24.98 \pm0.07}$ & $\boldsymbol{25.02 \pm0.08}$ & $\boldsymbol{25.05 \pm0.06}$ & $\boldsymbol{25.06 \pm0.03}$ & $\boldsymbol{25.13 \pm0.04}$ & $\boldsymbol{24.98 \pm0.03}$ & $\boldsymbol{0.05 \pm0.04}$ \\
\enddata
\end{deluxetable*}

\begin{figure*}
\centering
\includegraphics[width=0.85\textwidth]{"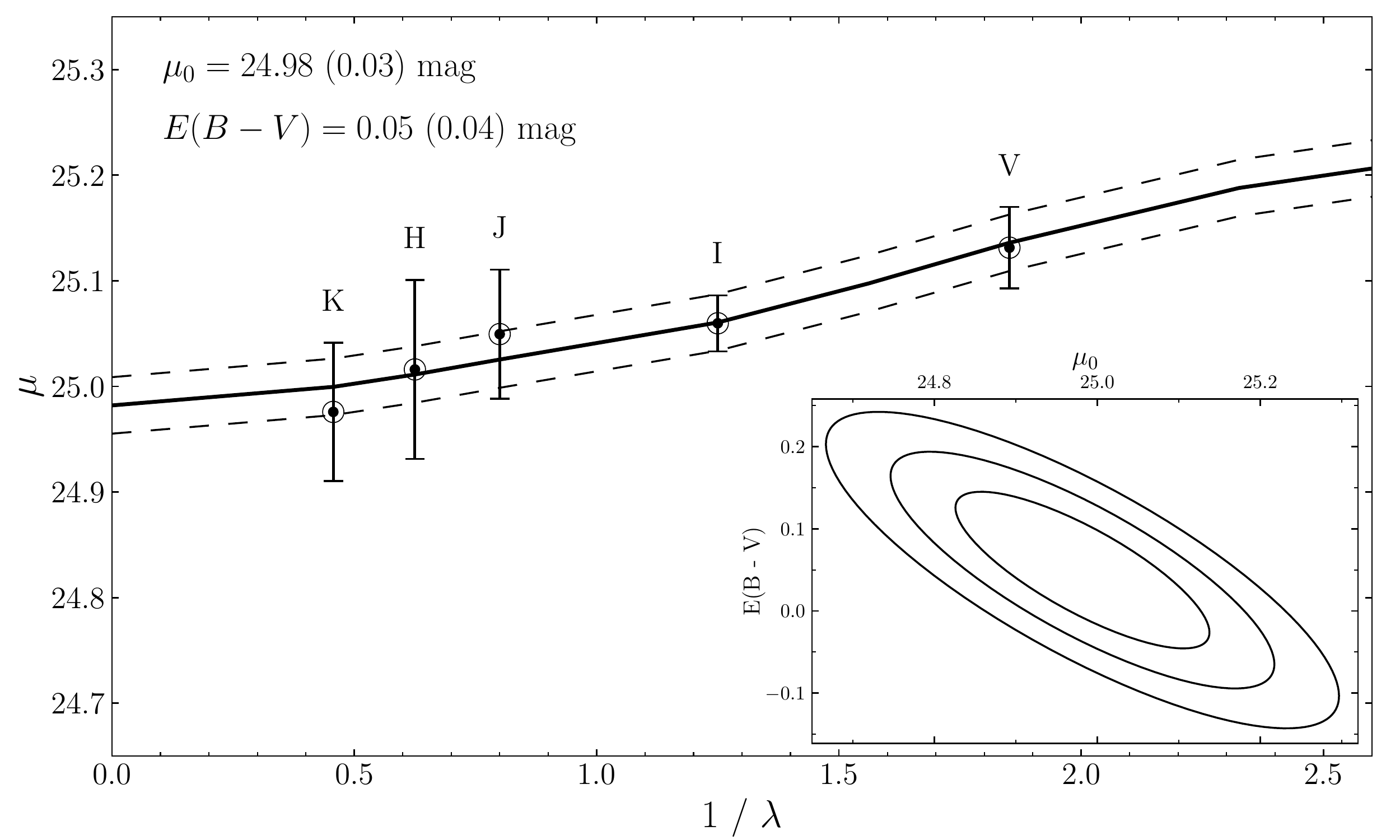"}
\caption{Extinction curve fit to the wavelength-dependent distance moduli derived from WLM Cepheids. Error bars on the individual points give the standard error on the mean as determined from the PL-relation. The solid line shows the best fit, and the dashed lines give the 1$\sigma$ error. The inset in the lower right corner shows the 1, 2, and 3$\sigma$ contours of the error space of the fit, with the true distance modulus on the x-axis and the reddening on the y-axis.
\label{fig:f9}}
\end{figure*}

Due to the relatively large errors on the individual distance moduli compared to the (minimized) scatter in the finally-adopted fit (see Figure \ref{fig:f9}), we chose to estimate the fit error from the data point uncertainties alone. We did this by assuming a Gaussian distribution for each error, with the width equal to the uncertainty on the point and the mean equal to the distance from the fit. We summed the distributions and calculated the plus and minus offsets from the mean that would each contain 34\% of the area. The offsets were equal in our chosen subsample, indicating a symmetrical distribution, and we determined a statistical error on the true distance modulus to be $\pm 0.03$ mag. Using this error value, the 1, 2, and 3$\sigma$ contours of the error space were calculated (see inset in Figure \ref{fig:f9}). We determined the $E(B-V)$ error by dividing half the full 1$\sigma$ contour range by the square root of the number of data points. This results in an error of $\pm0.04$ mag for our subsample.

The Cepheid Leavitt law distance modulus and reddening determined most recently by the Araucaria Group \citep{2008ApJ...683..611G} were $\mu_{0} = 24.92 \pm 0.04$ mag and $E(B-V) = 0.08 \pm 0.02$ mag. Our finally-adopted distance modulus agrees well with this earlier determination, with $\mu_{0} = 24.98 \pm 0.03$ (stat) $\pm 0.04$ (sys) mag. With this fit, we obtain a total line-of-sight reddening value of $E(B-V) = 0.05 \pm 0.04$ mag. Adopting an approximate value of 0.03 mag for the foreground galactic reddening to WLM \citep{2011ApJ...737..103S}, we obtain an intrinsic reddening of $0.02 \pm 0.04$ mag.

\begin{deluxetable*}{lclcc}
\tablenum{4}
\tablecaption{Previously Published TRGB and JAGB Distance Moduli to WLM\label{tab:history}}
\tablewidth{1pt}
\tablehead{
\colhead{Study} & 
\colhead{Filter \& Method} & 
\colhead{$\mu_{0}$ (published) [mag]} & 
\colhead{$\mu_{0}$ (standardized) [mag]} & 
\colhead{Notes} 
}
\startdata
\cite{1993ApJ...417..553L} &  $I_{~TRGB}$ & $24.81 \pm0.10$ & $24.85$\\
\cite{1997AJ....114..147M} & $I_{~TRGB}$ & $24.75\pm0.10$ & 24.80\\
\cite{1997MNRAS.289..406S} & $I_{~TRGB}$ & $24.97$ & $24.85$ & No errors given\\
\cite{2000ApJ...529..745F} & $I_{~TRGB}$ & $24.84$ & $-$ & No errors given, no original $I_{TRGB}$ given\\
\cite{2005MNRAS.356..979M} & $I_{~TRGB}$ & $24.85\pm0.08$ & $24.85$ \\
\cite{2007ApJ...661..815R} & $I_{~TRGB}$ & $24.93 \pm0.04$ & $24.92$\\
\cite{2009AJ....138..332J} & $I_{~TRGB}$ & $24.95^{+0.03}_{-0.09}$ & $24.95$\\
\cite{McCall} & $I_{~TRGB}$ & $24.82 \pm0.10$ & $24.85$\\
\cite{2017ApJ...834...78M} & $I_{~TRGB}$ & $24.94 \pm0.03$ & $24.91$\\
\cite{2019MNRAS.490.5538A} & $I_{~TRGB}$ & $24.93 \pm0.07$ & $24.91$\\
\hline
\cite{2011AJ....141..194G} & $J_{~TRGB}$ & $25.14\pm0.15$ & $-$ & No original $J_{~TRGB}$ given\\
 & $K_{~TRGB}$ & $25.12\pm0.17$ & $-$ & No original $K_{~TRGB}$ given\\
\hline
\cite{2020arXiv200510793F} & $JAGB$ & $24.97\pm0.05$ & $24.97$ \\
\hline
\enddata
\tablecomments{The $\mu_0$ ($TRGB_{I}$) has been standardized to $M_{I}^{TRGB}=-4.05$ mag from \cite{2020ApJ...891...57F} and an extinction of $A_I=0.057$ mag.}
\end{deluxetable*}

\subsection{Additional Distance Comparisons}
Finally, we compare our four measured distance moduli to the most precise previously published distances to WLM (obtained via NED) calculated with methods not used in this study. These measurements are based on the Horizontal Branch (HB) feature and CMD fitting techniques.

\cite{2000AJ....120..801R} measured a distance modulus of $\mu_0=24.95 \pm0.13$ mag based on the $V$-band magnitude of the HB, which is in 1-$\sigma$ agreement with all four of our distance measurements. 

There have also been three measured distance moduli to WLM based on CMD-fitting analysis techniques \citep{1999ApJ...521..577H, 2000ApJ...531..804D, 2014ApJ...789..147W}. Averaging them yields a distance modulus of $\mu_0=24.83\pm0.11$ mag, which is $1-2$ $\sigma$ from all four of our measured distance moduli. 

\section{Summary and Conclusions}\label{sec:summary}
In this paper we have presented consistent measured distances to WLM using the optical TRGB, NIR TRGB, JAGB method, and $VIJHK$ observations of Cepheids, demonstrating promise for the JAGB method as a competitive supernovae type Ia calibrator. We have also showed the JAGB's future effectiveness as an independent cross-check for systematics with the Cepheid and TRGB methods at NIR wavelengths.

We determined the optical TRGB distance modulus, measured in the $HST$ $F814W$ filter to be $\mu_0 (TRGB_{F814W}) =24.93 \pm 0.02$ (stat) $\pm0.06$ (sys) mag. Our second distance modulus, determined using the NIR TRGB, was $\mu_0 (TRGB_{NIR})=24.98 \pm0.04$ (stat) $\pm0.07$ (sys) mag. Third, we calculated our value for the JAGB distance modulus to be $\mu_0$ (JAGB) $=24.97 \pm0.02$ (stat) $\pm 0.04$ (sys) mag. Finally, we calculated a Cepheid distance modulus of $\mu_0$ (Cepheid)$=24.98 \pm0.03 \text{ (stat)} \pm0.04 \text{ (sys) mag}$. These correspond to distances of $968 \pm27$ kpc, $991 \pm37$ kpc, $986 \pm21$ kpc, and 991 $\pm$ 23~kpc for the optical TRGB, NIR TRGB, JAGB, and Cepheid-based methods, respectively.

The consistency of the distance moduli calculated across the optical TRGB, near-infrared TRGB, JAGB methods, and the Cepheid Leavitt law adds to the growing evidence that  the JAGB method is an excellent distance indicator and has promising potential for future work. The JAGB method is accurate, as shown by its consistency with the TRGB and Cepheids in its distance modulus measurement. In addition, the JAGB method is also precise, with error bars comparable to those based on optical measurements of the TRGB and multi-wavelength measurements of Cepheids.
Finally, the JAGB method is simple; only the mean magnitude of the JAGB distribution needs to be measured. 

The agreement in the measured TRGB and Cepheid distances for WLM is consistent with that found for other nearby galaxies; however, the two methods increasingly disagree out to further distances  \citep{2019ApJ...882...34F}, indicating possible systematic errors in one or both methods.
In the future, it it will be helpful to have independent JAGB distances for galaxies at farther distances for which the agreements starts to break down.

The full potential of the JAGB method as a distance indicator is still to be fully realized. Future studies using \textit{HST} and eventually \textit{JWST} data are capable of measuring distances well beyond the Local Group. 
We note that JAGB stars are about one mag brighter than the TRGB in the infrared (see Figure \ref{fig:f5}). Although long-period Cepheids  and Mira variables are brighter, they need increasingly longer baselines in order to secure their periods, ranging from 100 to up to 1000 days or more, which can become observationally expensive.
Our expectation is that JAGB stars will be able to calibrate type Ia supernovae hosts out to 100 Mpc.

With future work and development and a large sample of JAGB distances to supernovae type Ia hosts, the JAGB method will eventually result in its own determination of the Hubble Constant using near-infrared data from \textit{HST} and \textit{JWST}. Furthermore, the JAGB method can be used in tandem with the NIR TRGB and Cepheid-based methods on galaxies with low-reddening and low surface-brightness such as WLM; as shown in this study, the two methods can be used as a cross-checks and comparison for each other for revealing systematic errors.

\acknowledgments
We thank Peter Stetson for providing us with his copies of \textsc{daophot}, \textsc{daogrow}, and \textsc{allframe}. We  thank Violet Mager for providing useful comments on flat-fielding IMACS data. 
We thank Dylan Hatt for his help with the preparation of a figure. Finally, we thank the anonymous referee for their constructive suggestions.

Some of the data presented in this paper were obtained from the Mikulski Archive for Space Telescopes (MAST). STScI is operated by the Association of Universities for Research in Astronomy, Inc., under NASA contract NAS 5-26555. This research has made use of the NASA/IPAC Extragalactic Database (NED) and the NASA/IPAC infrared Science Archive (IRSA), both of which are operated by the Jet Propulsion Laboratory, California Institute of Technology, under contract with the National Aeronautics and Space Administration. 

This paper includes data gathered with the 6.5 meter Baade/Magellan Telescope located at Las Campanas Observatory, Chile.

This publication makes use of data products from the Two Micron All Sky Survey, which is a joint project of the University of Massachusetts and the Infrared Processing and Analysis Center/California Institute of Technology, funded by the National Aeronautics and Space Administration and the National Science Foundation.

\facility{HST (ACS/WFC), Magellan-Baade (IMACS, FourStar)}
\software{\textsc{daophot} \citep{1987PASP...99..191S}, \textsc{allframe} \citep{1994PASP..106..250S}, \textsc{daogrow} \citep{1990PASP..102..932S},
TinyTim \citep{2011SPIE.8127E..0JK}, \textsc{topcat} \citep{2005ASPC..347...29T}, Astropy \citep{2013A&A...558A..33A, 2018AJ....156..123A}, NumPy \citep{2011CSE....13b..22V}, Matplotlib \citep{2007CSE.....9...90H}, scipy \citep{2020NatMe..17..261V} }

\appendix 
\section{Photometric Quality Cuts}\label{App:qualitycut}
\begin{figure}
\figurenum{A1}
\gridline{\fig{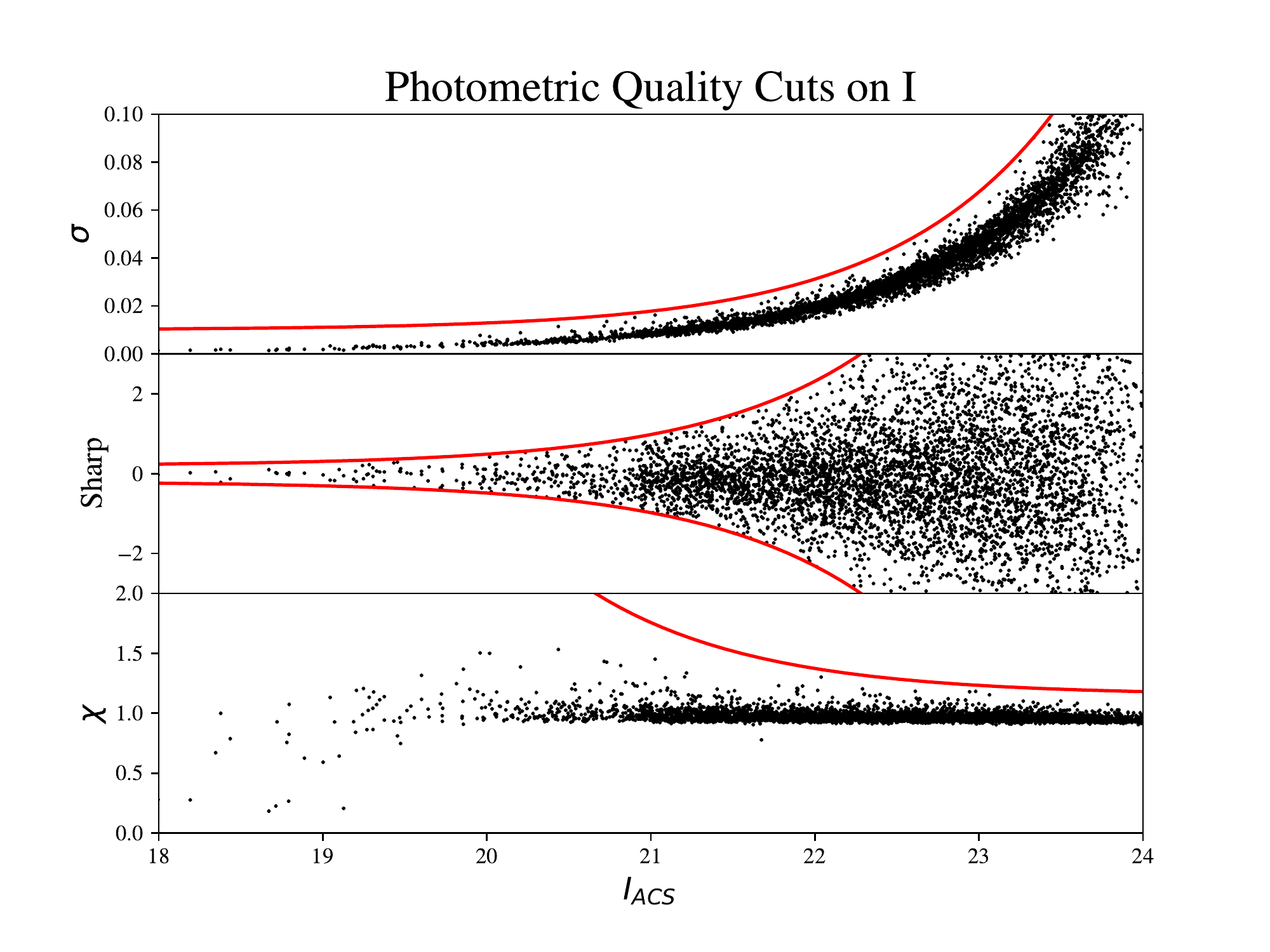}{0.5\textwidth}{}
          \fig{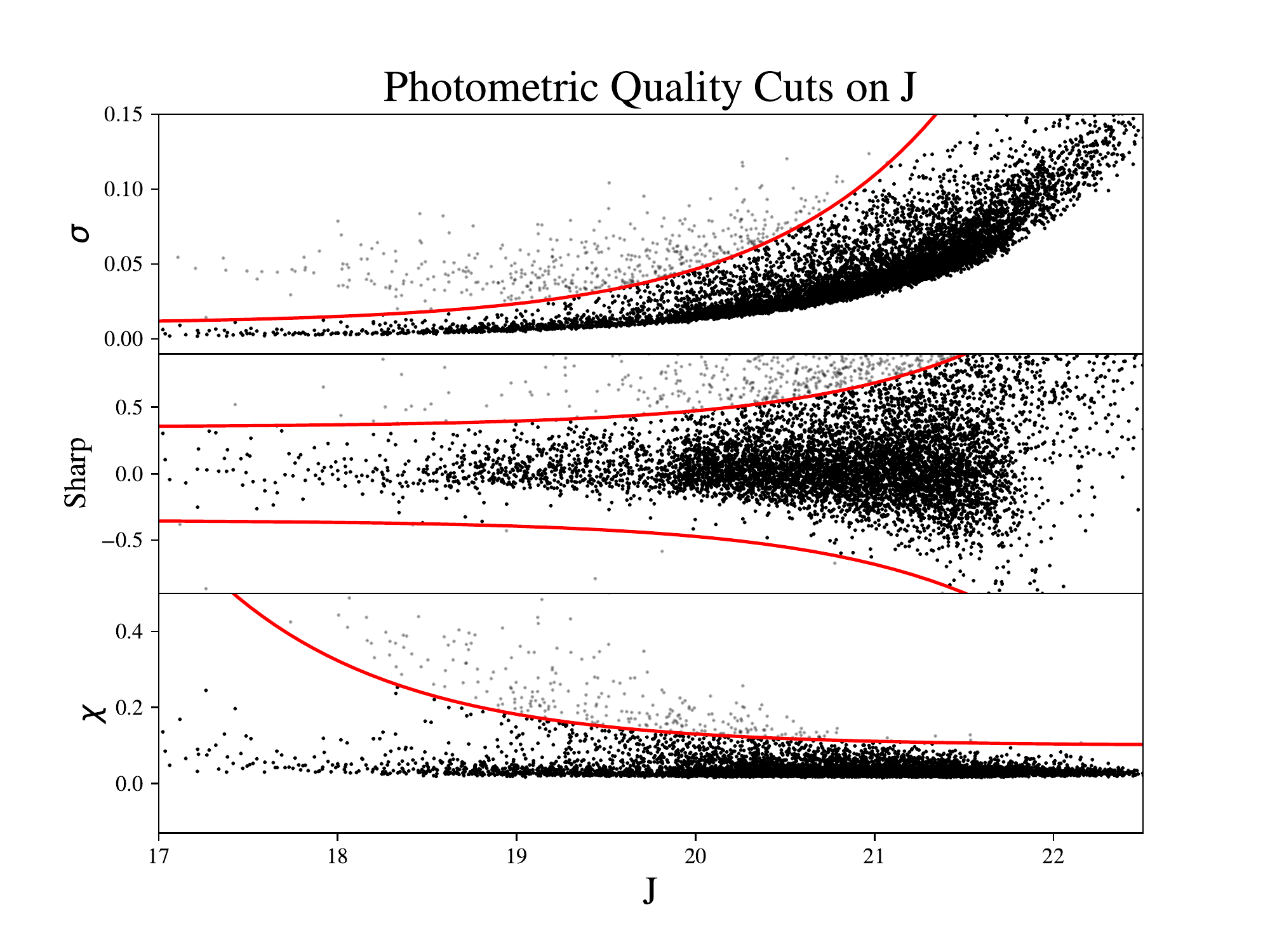}{0.5\textwidth}{}}
\gridline{\fig{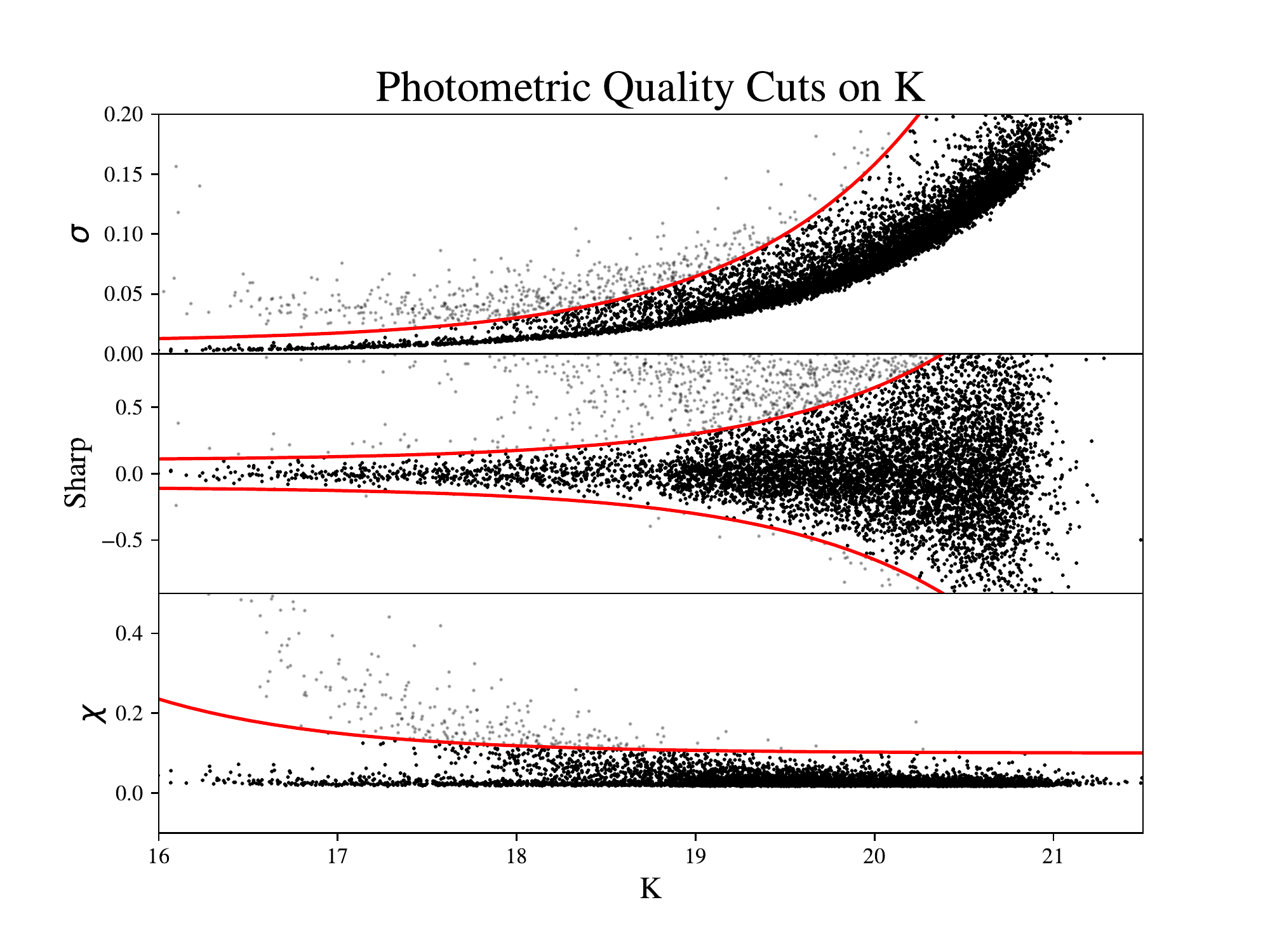}{0.5\textwidth}{}
          \fig{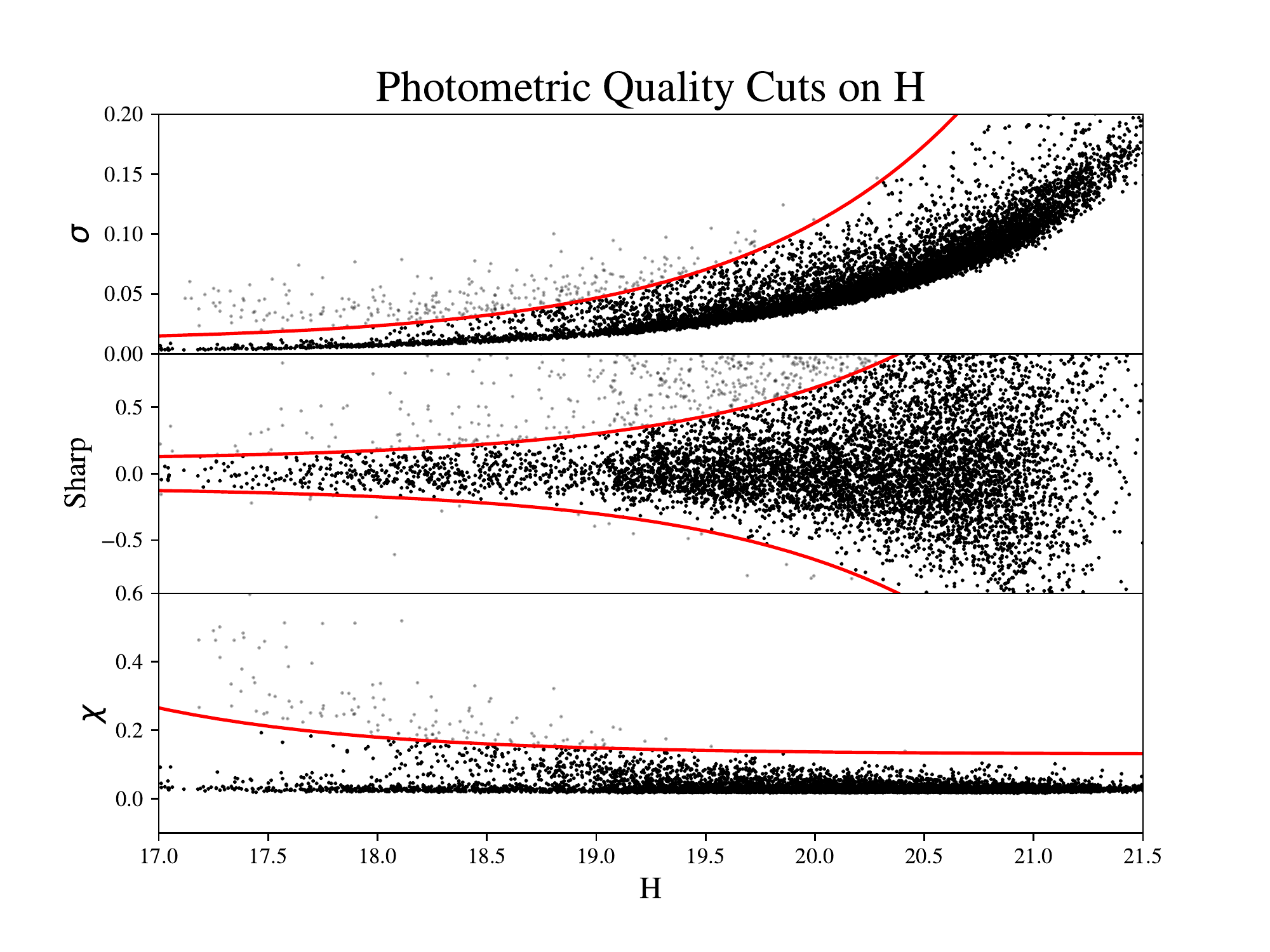}{0.5\textwidth}{}}
\caption{Photometric quality cuts made on $IJHK$ data. For the $JHK$ data, a source had to pass all three $\sigma, sharp$, and $\chi$ cuts for all three bands to be considered a stellar source. 
\label{fig:photcuts}}
\end{figure}

In order to limit our photometry to stellar sources and eliminate extended sources, we applied cuts on the following photometry parameters provided by \textsc{daophot}: the photometric uncertainty $\sigma$, the sharpness parameter, and the chi parameter $\chi$. To be considered a stellar source, every source had to pass all three cuts. In Figure \ref{fig:photcuts}, the black points show sources that passed all three cuts, and the grey points are sources that failed one or more of the cuts. We made cuts based on the following exponential functions as function of magnitude $m$, where the chosen values for each given function and filter can be found in Table \ref{tab:values}:

\begin{equation}
\sigma < a + b \times e^{m-c}
\end{equation}
\begin{equation}
|sharp| <a + b \times e^{m-c}
\end{equation}
\begin{equation}
\chi < a + e^{-(m-b)}
\end{equation}

\begin{deluxetable}{cccc}[h]
\tablenum{5}
\tablecaption{Photometric Quality Cuts \label{tab:cuts}}
\tablehead{
\colhead{Filter} & 
\colhead{$\sigma$ (a,b,c)} & 
\colhead{$sharp$ (a,b,c)} & 
\colhead{$\chi$ (a,b)} 
\label{tab:values}
}
\startdata
$I$ & 0.01,0.01,21.25 & 0.20, 70.0, 25.5 & 1.15, 20.5\\
$J$ & 0.01,0.003,17.5 & 0.15,0.01,15.5 & .35, 17.5\\
$K$ & 0.01,0.003,16.1 & 0.1,0.01,14.5 & .1,16\\
$H$ & 0.01,0.003,16.5 & 0.1,0.01,16 & .13,15
\enddata
\end{deluxetable}

\end{document}